\documentstyle[epsfig]{mn}

\def\etal{{\rm et al.\thinspace}}
\def\eg{{\it e.g.\ }}

\def\ie{{\it i.e.\ }}
\def\cf{{\it c.f.\ }}
\def\hi{H{\sc i} \,}

\def\nH{\hbox{$N_{\rm H}$}}
\def\Mdot{\hbox{$\dot M$}}
\def\h50{\hbox{$h_{50}$\,}}
\def\spose#1{\hbox to 0pt{#1\hss}}
\def\ltsimm{\mathrel{\spose{\lower 3pt\hbox{$\sim$}}
	\raise 2.0pt\hbox{$<$}}}
\def\ltsim{$\mathrel{\spose{\lower 3pt\hbox{$\sim$}}
	\raise 2.0pt\hbox{$<$}}$}
\def\gtsimm{\mathrel{\spose{\lower 3pt\hbox{$\sim$}}
	\raise 2.0pt\hbox{$>$}}}
\def\gtsim{$\mathrel{\spose{\lower 3pt\hbox{$\sim$}}
	\raise 2.0pt\hbox{$>$}}$}

\def\fract#1/#2{\leavevmode\kern.1em                   
   \raise.5ex\hbox{\the\scriptfont0 #1}\kern-.1em
   /\kern-.15em\lower.25ex\hbox{\the\scriptfont0 #2}}
\def\cm{{\rm\thinspace cm}}
\def\erg{{\rm\thinspace erg}}

\def\K{{\rm\thinspace K}}
\def\keV{{\rm\thinspace keV}}
\def\km{{\rm\thinspace km}}
\def\kpc{{\rm\thinspace kpc}}
\def\Lsol{\hbox{$\thinspace L_{\odot}$}}

\def\Msol{\hbox{$\thinspace M_{\odot}$}}
\def\pc{{\rm\thinspace pc}}
\def\s{{\rm\thinspace s}}

\def\yr{{\rm\thinspace yr}}

\def\ps{\hbox{\s$^{-1}\,$}}
\def\pcc{\hbox{$\cm^{-3}\,$}}
\def\pyr{\hbox{$\yr^{-1}\,$}}

\def\pcm2{\hbox{$\cm^{-2}\,$}}
\def\ergpcm3ps{\hbox{$\erg\cm^{-3}\s^{-1}\,$}}
\def\ergps{\hbox{$\erg\s^{-1}\,$}}

\def\kmps{\hbox{$\km\s^{-1}\,$}}

\def\Lsolppc3{\hbox{$\Lsol\pc^{-3}\,$}}
\def\Msolppc3{\hbox{$\Msol\pc^{-3}\,$}}

\title[Predicting X-ray emission from wind-blown bubbles]
{Predicting X-ray emission from wind-blown bubbles
 -- Limitations of fits to {\it ROSAT} spectra}
\author[David K. Strickland and Ian R. Stevens]
{David K. Strickland and Ian R. Stevens \\  
School of Physics and Astronomy, The University of
Birmingham, Edgbaston, Birmingham, U.K. B15 2TT \\
dks@star.sr.bham.ac.uk, irs@star.sr.bham.ac.uk} 

\date{Accepted .....; Received .....; in original form .....}

\pagerange{\pageref{firstpage}--\pageref{lastpage}} \pubyear{1998}

\begin{document}

\maketitle
\label{firstpage}

\begin{abstract}
Wind-blown bubbles, from those around massive O and Wolf-Rayet stars, 
to superbubbles around OB associations
and galactic winds in starburst galaxies, have a dominant role in
determining the structure of the Interstellar Medium. X-ray observations
of these bubbles are particularly important as most of their volume is
taken up with hot gas, $10^{5} \ltsimm T (\K) \ltsimm 10^{8}$. 

However, it is difficult to compare these X-ray observations, usually
analysed in terms of
single or two temperature spectral model fits, with theoretical models,
as real bubbles do not have such simple temperature 
distributions. Spectral fits, and the properties inferred from them,
will depend in a complex way on the true temperature distribution and
the characteristics and limitations of the X-ray observatory used.

In this introduction to a series of papers detailing the {\em observable}
X-ray properties of wind-blown bubbles, we describe our method with which we 
aim to solve this problem, analysing a simulation of a wind-blown
bubble around a massive star.

Our model is of a wind of constant mass and energy injection rate,
blowing into a uniform ISM, from which we calculate X-ray spectra as would be
seen by the {\it ROSAT} PSPC. Analysing these spectra in the same way
as a real observation would be, we compare the properties of the bubble
as would be inferred from the {\it ROSAT} data with the true properties of the
bubble in the simulation.

We find standard spectral models yield inferred properties that deviate 
significantly from the true properties, even though the spectral fits are
statistically acceptable, and give no indication that they do not
represent to true spectral distribution.
For example, single temperature spectral fits give best fit metal abundances
only 4\% of the true value. A cool bubble has best fit temperatures
significantly higher than a bubble twice as hot.
These results suggest that in any case where
the true source spectrum does not come from a simple single or two
temperature distribution the ``observed'' properties cannot naively be
used to infer the true properties. In this situation, to compare X-ray 
observations with theory it is necessary to calculate the 
{\em observable} X-ray properties of the model.
\end{abstract}


\section{Introduction}
Bubbles blown in the Interstellar Medium (ISM) by massive stars are a
common astrophysical phenomenon. X-ray observations can provide
information regarding the density, metal abundance, temperature,
ionisation state and physical structure in the hot bubbles surrounding
Wolf-Rayet (WR) and O stars (Wrigge, Wendker \& Wisotski 1994),
Luminous Blue Variables (LBV's) such as $\eta$ Carinae (Corcoran \etal
1995) and planetary nebulae (PN) (Kreysing \etal 1992; Leahy, Zhang \&
Kwok 1994; Arnaud, Borkowski \& Harrington 1996; Leahy \etal 1996). On
the larger scale, superbubbles are created by the sum of the winds and
SN within OB associations (Belloni \& Mereghetti 1994) and giant star
forming regions in young starburst galaxies.  Superbubbles within
starburst galaxies such as M82 eventually break out the galaxy to form
spectacular galactic winds (Watson, Stanger \& Griffiths 1984;
Heckman, Armus \& Miley 1987). In many cases the X-ray emission
probes different regions of the object in question to that revealed by
optical observations, increasing the importance of the X-ray data.

Analytic solutions to the development and structure of astrophysical
bubbles must rely on simplifying assumptions, and increasingly
attention has turned to the use of numerical hydrodynamics.
These simulations have been enlightening with respect
to the nonlinear processes occurring, with some degree of quantitative
agreement with observation, but generally lack predictive power. This
is partially due to the difficulty in comparing them with observations,
in particular X-ray observations.

The problem is that X-ray observations are usually analysed by fitting
a single or two temperature spectral model to the observed spectra
 (see for example the references above),
and the best-fit results are used to infer the physical properties
of the object.

However, for wind-blown bubbles such as those mentioned above, the true
situation is more complex, and the results of the spectral fits may
be influenced by, for example:
projection of different physical regions
along the line of sight; the presence of a wide range of temperatures;
interstellar absorption; unknown or non-standard elemental abundances;
non-ionisation equilibrium conditions;
low numbers of observed photons and the limitations of the current
X-ray telescope optics and detectors. 

All of these make the
interpretation of what is normally a one or two temperature spectral
fit to the data difficult to relate to the underlying physical
conditions, and conversely, make it difficult to predict the
{\em observable} properties of a model or simulation.

To our knowledge there has been no study of the influence of the
complexities mentioned above on the best-fit properties of a spectral fit
to the observable X-ray data, 
and in particular not for wind blown bubbles. As we shall show,
the combination of the physical effects above and the properties 
(and limitations) of real X-ray observatories, can significantly
affect the results of simple spectral fitting.

Previous authors (Weaver \etal 1977;  Zhekov \& Perinotto 1996)
have calculated theoretical X-ray spectra from their 1-D models,
but did not consider particular instruments or fit models to those
spectra. In general only X-ray luminosities are calculated 
(\eg Volk \& Kwok 1985; Mellema \& Frank 1995; 
Garcia-Segura \& Mac Low 1995).

The aim of this paper is to introduce a method of analysing numerical
simulations in the same way as actual X-ray observations are
analysed, \ie predict the {\em observable} X-ray properties. 
This method can be applied to a wide range of phenomenon
where X-rays are important, from wind-blown bubbles around WR stars
and PN, through the larger bubbles around clusters of massive stars to
starburst-driven galactic winds.

We simulate a wind blown bubble using a 2-dimensional hydrodynamic
code, concentrating on the properties of the hot X-ray emitting
gas. The hydrodynamic model is used to generate artificial X-ray
spectra and images, in particular simulated {\it ROSAT} spectra. We
then analyse these spectra in the same way as real {\it ROSAT} spectra
would be, in order to determine what the observationally determined
properties of the bubble would be, and how those relate to its actual
structure.  This synthesis is necessary to a) determine the physical
processes that are observationally important, and b) {\em allow a direct
comparison between observation and theory}.

Our model of a wind-blown bubble is deliberately chosen to be the
simplest applicable model with an analytic solution, in order to 
simplify the analysis of our results, and avoiding 
added complications that a more physically accurate model of a wind
blown bubble (\eg Garc\'{\i}a-Segura, Mac Low \& Langer 1996)  would
introduce into the interpretation of our results. Later papers
will consider more realistic models, with additional physics such as
time varying energy and mass injection rates, along with spatial
variation of the X-ray properties. This will be necessary to
understand the properties of the extended emission from galactic winds
(see for example Strickland, Stevens and Ponman 1997).

In Section~\ref{sec:num_method} we describe the numerical code used to
produce the results shown in
Section~\ref{sec:results}. Section~\ref{sec:disc}  discusses the
implications of these results, and we briefly sum up in
Section~\ref{sec:conclusions}.

\begin{figure*}
\vspace{16.0cm}
\includegraphics{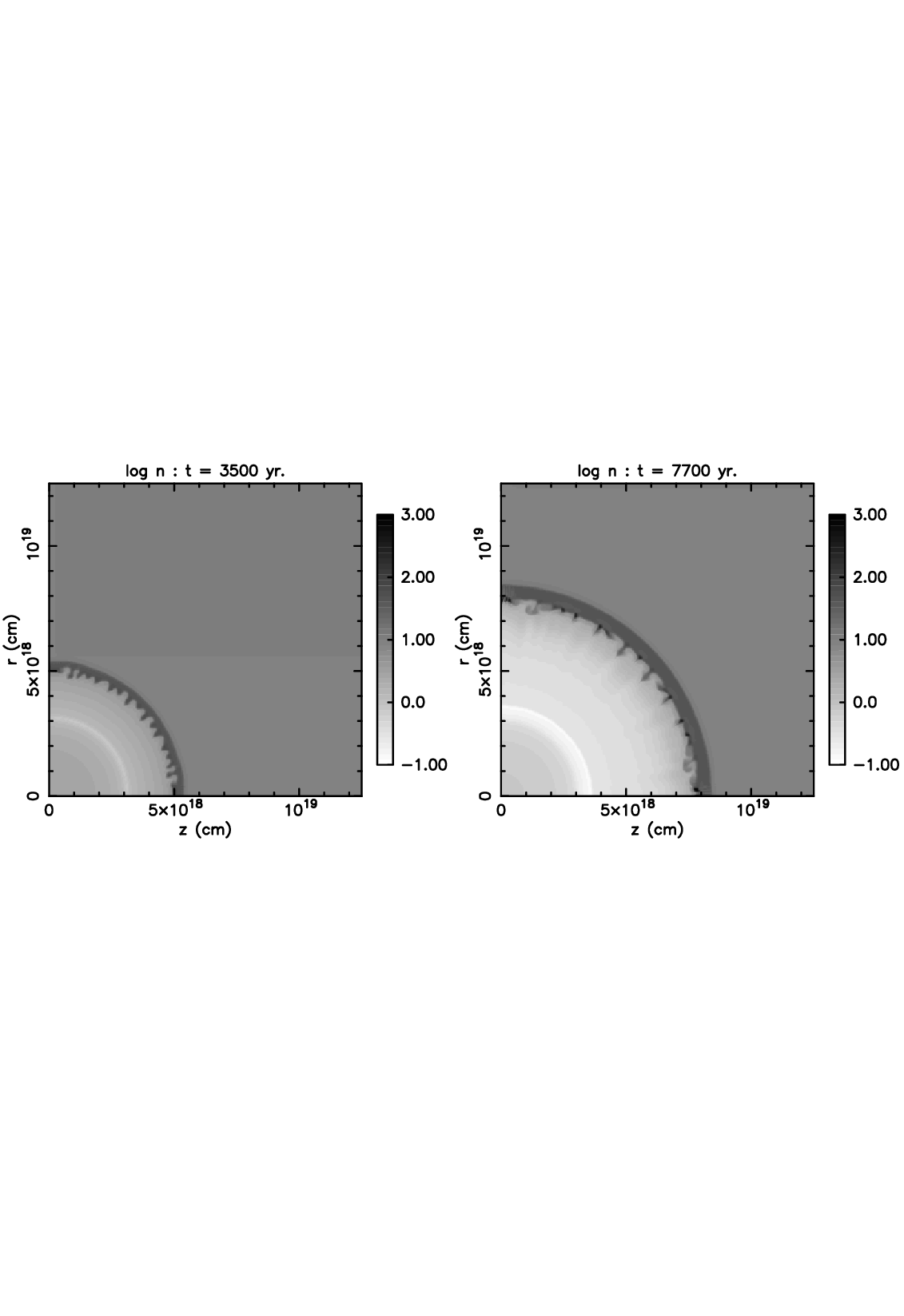}
\includegraphics{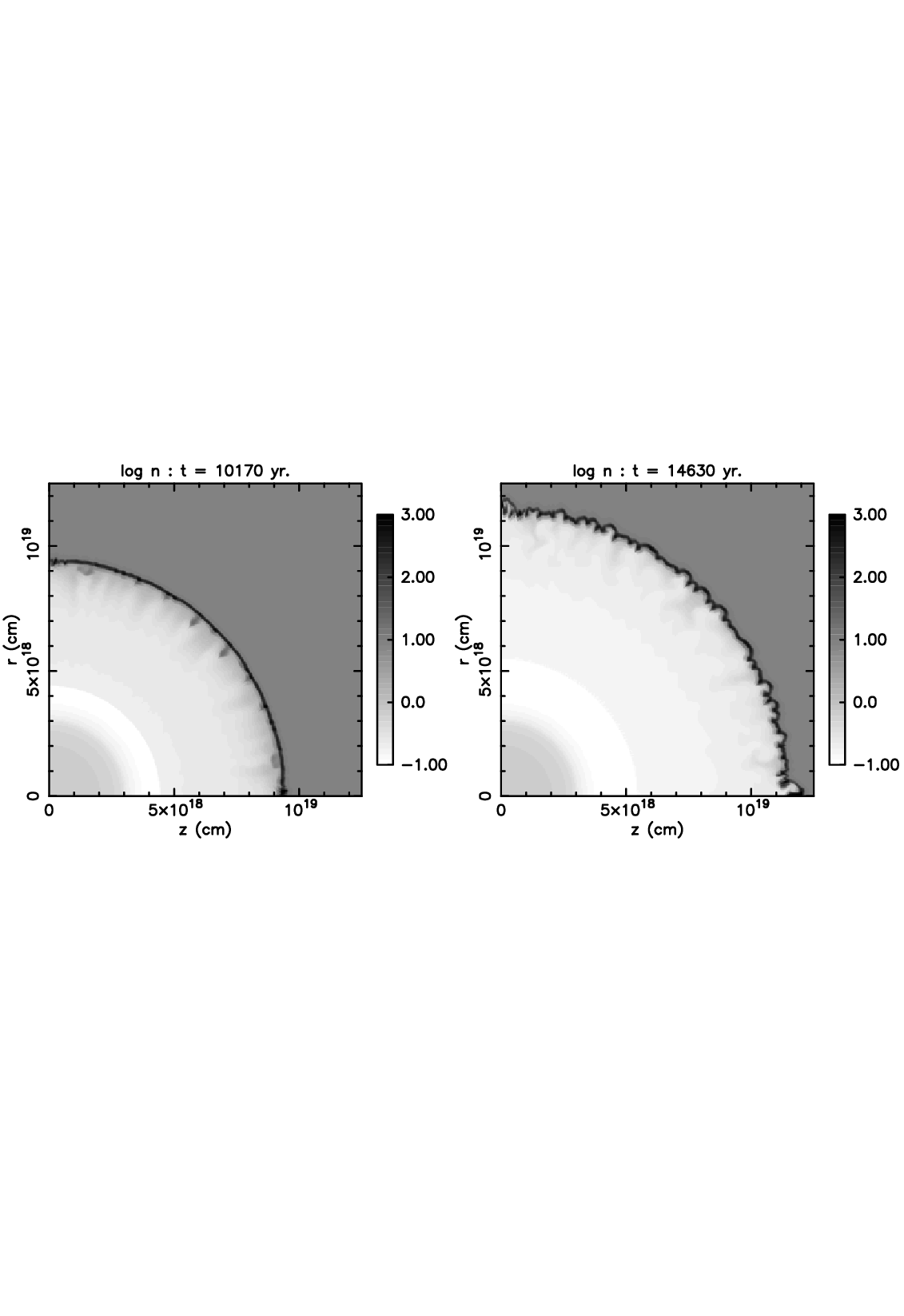}
\caption{Logarithm of the gas number density during the simulation at
$t = 3500$, $7700$, $10170$ and $14630 \yr$. At $t = 3500 \yr$ the
bubble has suffered no significant radiative energy loss. Shell
collapse is underway at $7700 \yr$, approximately the time of maximum
soft X-ray luminosity, and has just finished at $10170 \yr$. The
bubble then enters the self-similar phase, its properties at $t =
14630 \yr$ being typical of this stage.}
\label{fig:dens_4t}
\end{figure*}

\section{Numerical method}
\label{sec:num_method}
The 2-D numerical simulations presented in this paper were performed
using {\sc Virginia Hydrodynamics-1} ({\sc VH-1}), written by John
Blondin and co-workers (Blondin \etal 1990). VH-1 is based on the
piecewise parabolic method (PPM) of Colella \& Woodward (1984), which
is a third order accurate finite difference algorithm. For a more
detailed description of the code see Stevens, Blondin \& Pollock
(1992).

Radiative cooling is handled assuming optically thin gas with solar
abundance ratios and ionisation equilibrium. The cooling curve for the
temperature range $4.0 \leq \log T \leq 8.6$ was generated using the
Raymond-Smith plasma code (Raymond \& Smith 1977), assuming solar
element abundances. Because cooling times at $T \sim 10^{5} \K$ are
short, we place a limit of the numerical timestep to prevent more than
a $2\%$ energy loss per computational cell per timestep. For $T \leq
10^{4} \K$ the cooling is set to zero.

For the simulations described below VH-1 was run in cylindrical
coordinates ($r$, $z$), assuming symmetry around the $z$-axis. In
practice this gives rise to numerical artifacts along the coordinate
axes, but these are small and do not affect our results.  The
computational grid consisted of $400 \times 400$ cells of equal size,
spanning a physical distance of $r_{\rm max} = z_{\rm max} = 2.0\times
10^{19} \cm$.

\subsection{Simple wind-blown bubble model}
\label{sec:simp_mod}
As a simple realistic model of a wind-blown bubble we consider the
analytical model of Castor, McCray \& Weaver (1975) and Weaver \etal
(1977). A wind of constant terminal velocity $v_{\rm W}$ and
mechanical luminosity $L_{\rm W}=\frac{1}{2} \Mdot_{\rm W} v_{\rm
W}^{2}$ (where $\Mdot_{\rm W}$ is the mass loss rate of the star)
blows into a medium of uniform density $\rho_{0}$. This generates a
bubble with a structure divided into three distinct regions: an outer
shock, separating the undisturbed ambient ISM from a shell of swept up
and shocked ISM; a bubble of hot shocked stellar wind material,
bounded at its outer edge by a contact discontinuity between it an the
shell of shocked ISM, and at its inner edge by a reverse shock between
it and the third innermost region of the freely expanding stellar
wind. After several thousand years the shell of swept up ISM cools and
collapses down to form denser, cold ($T\sim 10^{4} \K$) shell, which
is the source of the the optical emission associated with the nebula.

In the standard Weaver \etal (1977) model thermal conduction leads to
evaporation of matter off the dense shell of swept-up ISM into the hot
bubble interior, cooling it and significantly increasing its
density. In common with many of the hydrodynamical simulations of
bubbles we do not include conduction. Magnetic fields may suppress
conduction, even at very low B-field strengths that otherwise are
dynamically unimportant (Soker 1994; Band \& Liang 1988). Given the
limited observational knowledge on the state of magnetic fields and
conduction within bubbles we choose to ignore both!

We shall consider two simulations, both with 
$L_{\rm W} = 6.3 \times 10^{37} \ergps$. To obtain different temperatures
in the hot bubble we vary the wind mass loss rate between simulations.
The low mass loss rate simulation has 
$\Mdot_{\rm W} = 5 \times 10^{-5} \Msol \pyr$. In the high mass loss rate 
simulation $\Mdot_{\rm W} = 1 \times 10^{-4} \Msol \pyr$, giving a bubble
of half the temperature of the low mass loss rate simulation. 
Mass and energy are
added to cells within $r = 3 \times 10^{18} \cm$ at each timestep.
The ambient medium is assumed to be uniform and totally ionised, with
a total number density of $n_{0}= 10 \pcc$. 


\begin{table}
\caption{Model parameters for the two bubble simulations}
\label{tab:model_params}
\begin{tabular}{ll}
Parameter                        & Value \\  
Wind luminosity $ L_{\rm W} $    & $ 6.3 \times 10^{37} \ergps $ \\  
Mass loss rate $ \Mdot_{\rm W} $ & $ 5 \times 10^{-5}$ or
                                   $ 1 \times 10^{-4} \Msol \pyr $ \\
Wind velocity $ v_{\rm W} $      & $ 2000$ or $1414 \kmps $ \\  
ISM ambient density $ n_{0} $    & $ 10 \pcc $ \\  
Distance to bubble               & $ 2 \kpc $ \\ 
Absorbing column $\nH$           & $ 3.16 \times 10^{21} \pcm2 $ \\
Grid size                        & $ 2.~ 10^{19} \cm \times 2.~10^{19} \cm $\\
Cell size                      & $ 5.~ 10^{16} \cm \times  5.~ 10^{16} \cm $ \\
\end{tabular}
\end{table}

\begin{figure*}
\vspace{16.0cm} 
\includegraphics{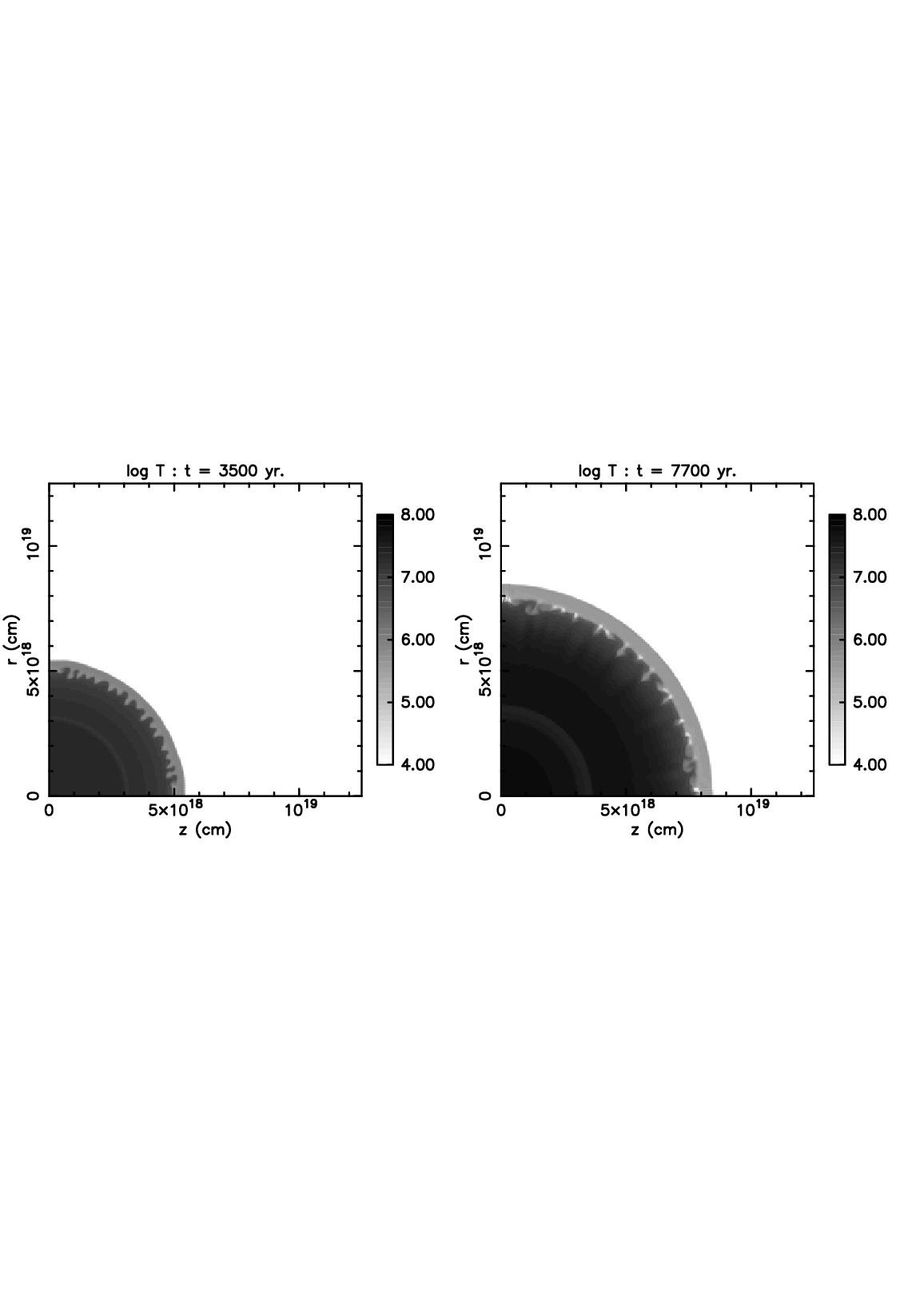}
\includegraphics{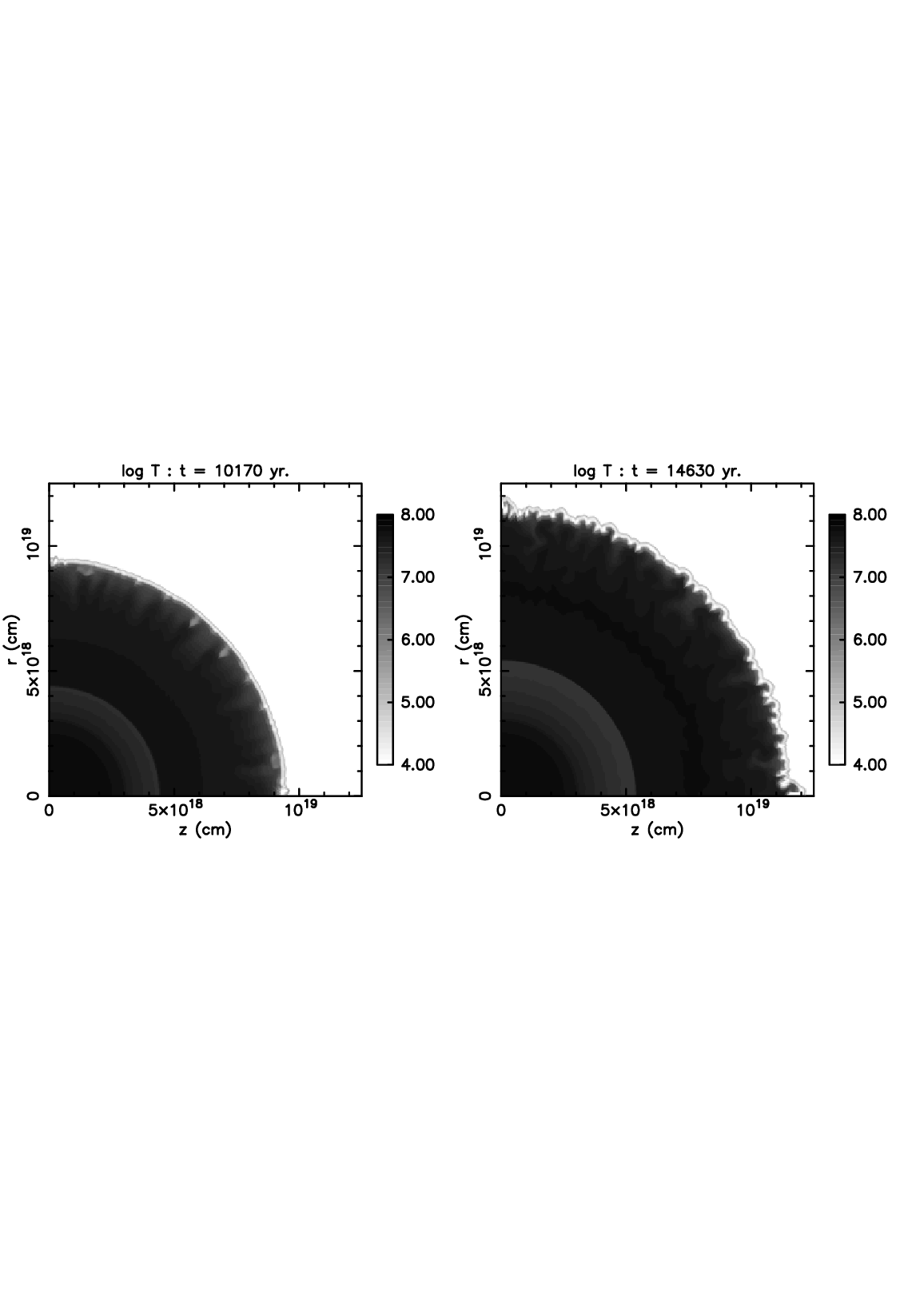}
\caption{As Fig.~\ref{fig:dens_4t} but greyscales of $\log_{10} T (\K).$}
\label{fig:temp_4t}
\end{figure*}

\subsection{Simulated X-ray data}
We shall only consider simulated {\it ROSAT} PSPC (Position Sensitive
Proportional Counter) data. Although the gas proportional counter's
spectral resolution of $\Delta E/E \approx 0.43 (E/0.93)^{-0.5}$
(FWHM, with E measured in $\keV$) is low compared to a mission such as
{\it ASCA}, wind-blown bubbles are soft X-ray sources and {\it ROSAT}
has more sensitivity than {\it ASCA} at low energies.

For a detailed discussion of the {\it ROSAT} satellite see `The {\it
ROSAT} user's handbook' (Briel \etal 1994).  



\section{Results}
\label{sec:results}
As the low and high mass loss rate simulations 
only differ in the density and temperature of the 
shocked wind, we shall concentrate on describing the low mass loss rate
simulation below. Section~\ref{sec:res_cool_bub} describes how the results
of the high mass loss rate bubble differ from those given below.

\subsection{Bubble growth}
There are three definable stages of bubble growth 
seen in the simulation. These
are i) before the shell cools, ii) during shell cooling and
collapse, and iii) self-similar growth after shell collapse 
with a thin cold ($T \sim 10^{4}
\K$) shell. The 1-dimensional analytic solutions for the first and
last of these stages are presented in detail in Castor \etal (1975)
and Weaver \etal (1977).

\begin{figure}
\vspace{9.0cm} 
\includegraphics{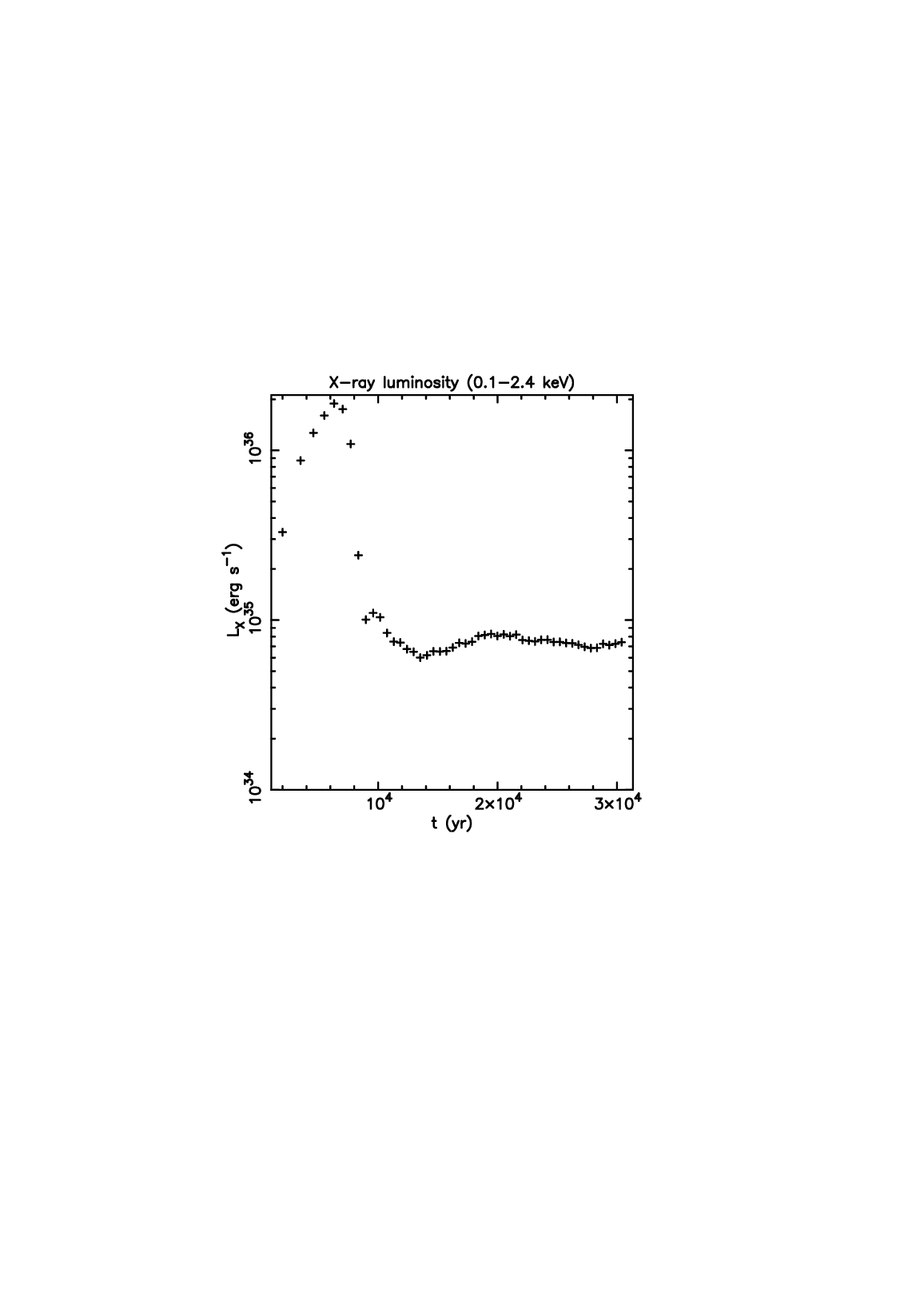}
\caption{Bubble X-ray luminosity as a function of time in the {\it
ROSAT} $0.1$-$2.4 \keV$ band.}
\label{fig:lx}
\end{figure}

\begin{figure*}
\vspace{16.0cm}  
\includegraphics{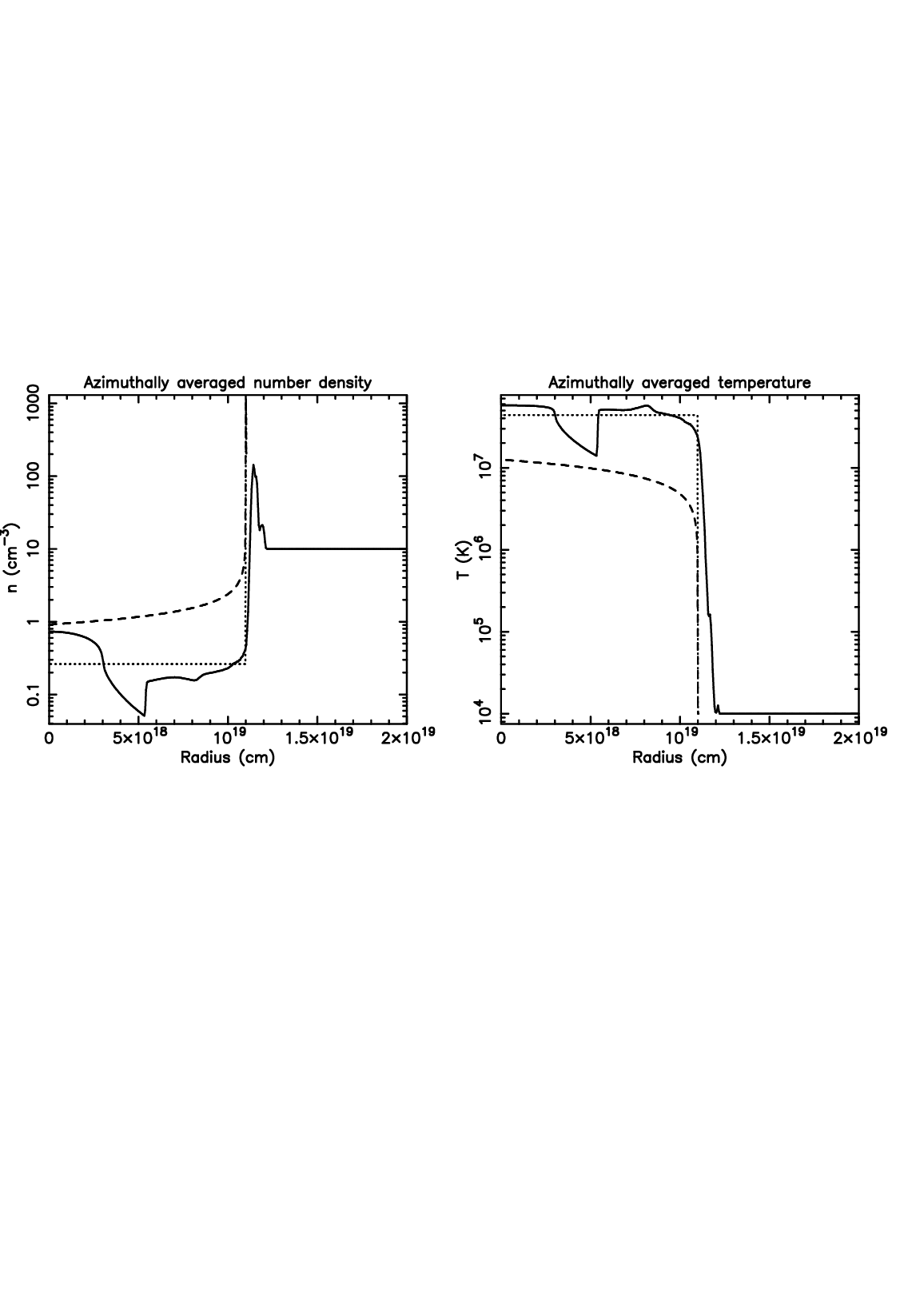}
\includegraphics{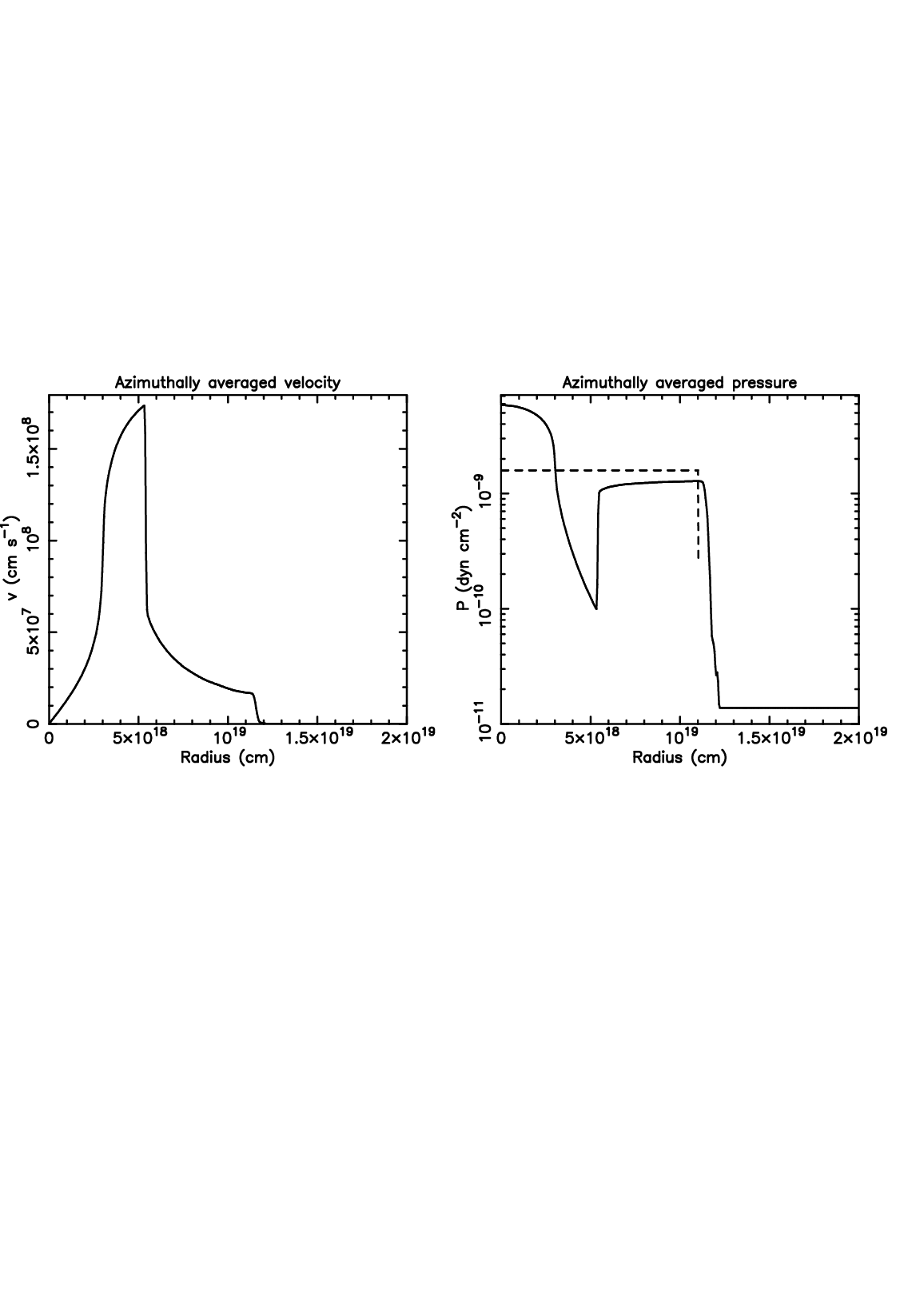}
\caption{Radial profiles of density, temperature, velocity
and pressure (solid lines) from the low mass loss rate simulation, 
compared to the predictions from of the
Weaver \etal (1977) model (with conduction: dashed line, no
conduction: dotted line), at $t = 14630 \yr$. Note that the Weaver model
only applies for $R > R_{1}$, the radius of the wind termination shock.}
\label{fig:weaver}
\end{figure*}

\begin{figure}
\vspace{9cm}
\includegraphics{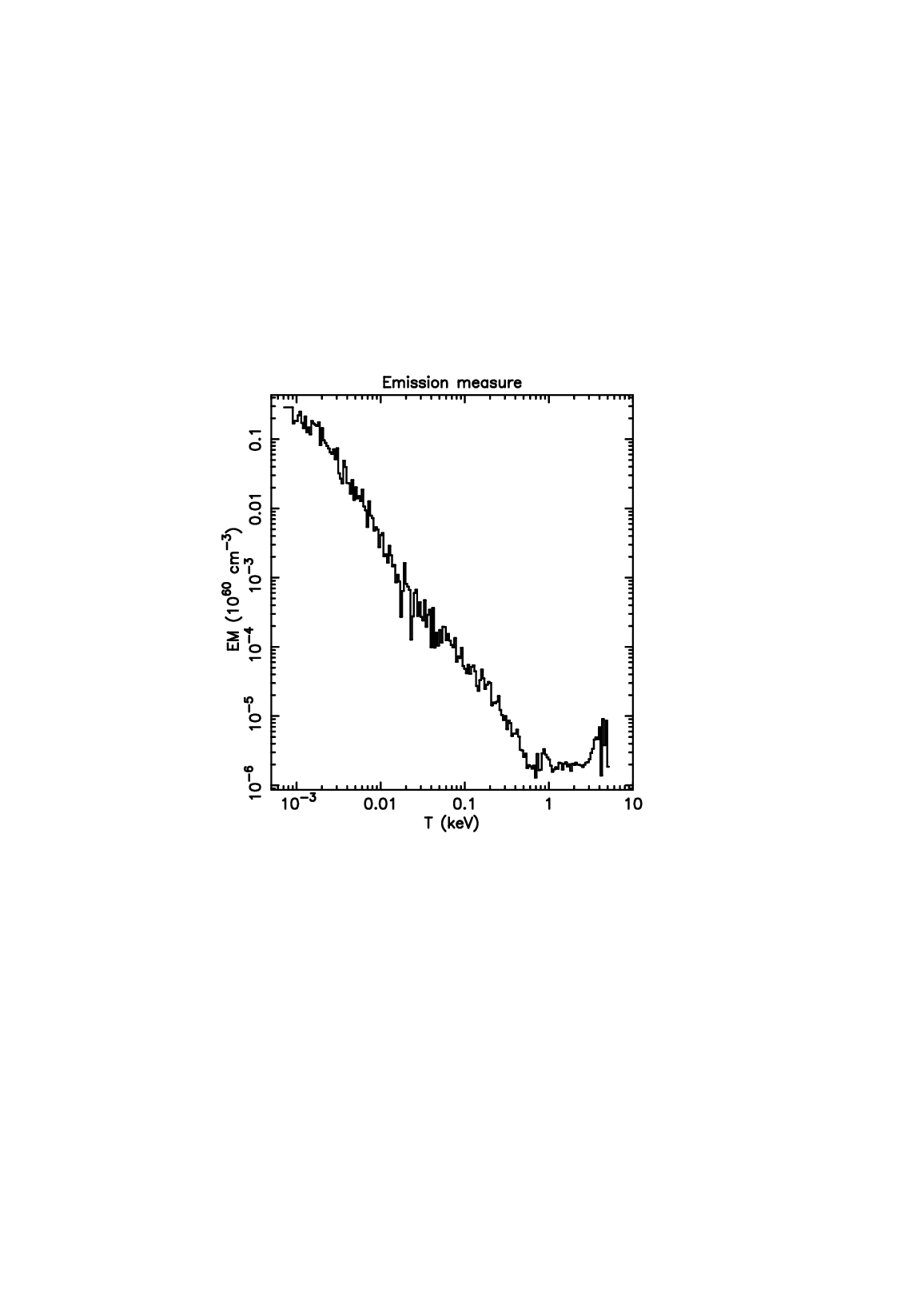}
\caption{The emission measure as a function of temperature ($EM_{T} =
\int n_{\rm e}^{2} \times dV_{T}$) for the entire bubble at $t = 15000 \yr$.}
\label{fig:em}
\end{figure}

Initially the swept up ISM is shock-heated to $5.5 \ltsimm \log T {\rm
(K)} \ltsimm 6.0 $.  The shell is thick (see
Figs.~\ref{fig:dens_4t} and \ref{fig:temp_4t}), and a strong emitter
of extreme Ultraviolet (EUV) radiation and soft X-rays, as can be seen
from the $0.1$-$2.4 \keV$ luminosity (Fig.~\ref{fig:lx}). The major
coolant is radiation in the UV-EUV rather than X-rays, the UV-EUV
luminosity being of order a magnitude greater than the soft X-ray
luminosity before shell collapse.  The luminosity rises rapidly with time, 
as the bubble sweeps up and heats more ISM. The rate of increase of
luminosity decreases after $t \approx 4000 \yr$, the X-ray luminosity
peaking at $L_{\rm X} = 1.9 \times 10^{36} \ergps$ at $t \approx 6300
\yr$ as the shell begins to cool. The UV-EUV luminosity peaks later,
at $\approx 8100 \yr$ as the shell cools further out of the X-ray band
and becomes denser. The peak UV-EUV luminosity of $9.1 \times 10^{37}
\ergps$ briefly exceeds the wind energy input. After shell collapse
X-ray luminosities are approximately two orders of magnitude below the
UV-EUV luminosity, both remaining essentially constant for the
duration of the simulation.

Following shell collapse shell densities are typically several hundred
to a few thousand particles per cubic centimetre, with $T \approx
10^{4} \K$.

In the absence of heat conduction and evaporation off the cool shell,
we would simplistically expect the bubble interior to have a uniform
low density and high temperature. In this case the shocked-wind
material has the same pressure as predicted by Castor \etal (1975) and
Weaver \etal (1977), but the temperature determined by the reverse
shock (the termination shock of the freely expanding wind).

In practice, the density rises in as the shell is approached, and the
temperature drops, although not to the extent expected for true
conduction. This can be seen in both the 2D images of
Figs.~\ref{fig:dens_4t}-\ref{fig:temp_4t} and the radial profiles of
Fig.~\ref{fig:weaver}. This is due to eddies and swirling motions
along the shell-bubble interface mixing material from the dense shell
into the hot bubble interior. The outward velocity in the shocked-wind
is higher than the velocity at which the shell expands into the ISM
(Fig.~\ref{fig:weaver}), and coupled with the corrugations seen on the
inside surface of the shell, shear motions arise between the faster
bubble interior and the shell, leading to swirling motions along the
interface and the introduction of cooler, denser material into the hot
bubble.

In a perfectly spherically symmetric bubble, the lack of a tangential
velocity component between the faster-expanding bubble interior and
the shell would prevent such stripping of material off the shell. In
our simulations, the bubble-shell interface is corrugated by
instabilities from early on in the simulation, presenting faces not
totally perpendicular to the flow in the bubble interior, and
leading to mixing.

\subsubsection{Shell instabilities}
\label{sec:results_instabs}
The instabilities of the shell seen in Figs.~\ref{fig:dens_4t} and
\ref{fig:temp_4t} have important consequences as they lead to the 
introduction of cooler,
denser material into the bubble interior, hence modifying the X-ray
emitting properties of the bubble.

The shell should be stable against Rayleigh-Taylor instabilities, as
it is constantly decelerating, suggesting that the instabilities
are Vishniac instabilities (Vishniac 1983).

The initial seed perturbation is numerical artifact, arising when the
forward shock first appears at the start of the simulation, and is 
due to the orthogonal nature of the computational grid and the finite size of
the energy injection region.

Does this instability, and the cooler material it injects into the
hot shocked wind region, render the results of these simulations
untenable?
In real system such as wind-blown bubbles around individual stars or 
superbubbles, both the wind energy injection 
rate and the ISM will be far from
uniform, so instabilities and inhomogeneities in the shell are to be
expected. Note that conduction would introduce far more cool
material into the bubble interior than these instabilities do
(Fig.~\ref{fig:weaver}). Mass loading of the flow by ablation of
clumps and cloudlets (\cf Hartquist \etal 1986) will also add
cool dense material.
We are therefor not worried that our 2D
simulations do not exactly match the 1D analytic prediction, as
the mass introduced into the bubble interior is by no means 
unphysically large.


\subsection{Bubble structure and properties at $\bmath{t=15 000 \yr}$}
\label{sec:res_t15000}
For a more detailed study of the observable X-ray properties we shall
investigate the bubble at $t = 14630 \yr$ (hereafter $15000 \yr$ for
convenience). The bubble at this time has settled down into
self-similar expansion with a thin shell, the period described by the
standard solution of Castor \etal(1975) and Weaver \etal (1977).

\begin{table}
\caption{Bubble properties at $t = 15 000 \yr$. The X-ray fluxes
in the {\it ROSAT} $0.1$-$2.4 \keV$ band and {\it ROSAT} PSPC
count rates are given for zero column (intrinsic) and $\nH = 3.16 \times
10^{21} \pcm2$ (absorbed).}
\label{tab:bub_properties}
\begin{tabular}{ll}
Parameter                     & Value \\  
Reverse shock radius $R_{1}$ & $ 5.3 \times 10^{18} \cm $ \\  
Forward shock radius $R_{2}$  & $ 1.17 \times 10^{19} \cm $ \\  
Total energy radiated         & $ 1.057 \times 10^{49} \erg $ \\  
Thermal energy                & $ 1.169 \times 10^{49} \erg $ \\  
Kinetic energy                & $ 6.824 \times 10^{48} \erg $ \\ 
Shell velocity                & $126$-$166 \kmps$ \\  
Total volume                  & $ 6.702 \times 10^{57} \cm^{3} $ \\  
Total mass                    & $ 34.4 \Msol $ \\  
Total emission measure        & $ 4.409 \times 10^{60} \pcc $ \\ 
Luminosity (0.1-2.4 $\keV$)   & $ 6.55 \times 10^{34} \ergps $ \\  
$f_{X}$ (int.)                & $ 1.36 \times 10^{-10} \ergps \pcm2 $ \\ 
PSPC count rate               & $ 35.3$ photons $\ps$ \\  
$f_{X}$ (abs.)                & $ 3.54 \times 10^{-12} \ergps \pcm2 $ \\  
PSPC count rate               & $ 0.3156$ photons $\ps$ \\
\end{tabular}
\end{table}

\subsubsection{Comparison with the Weaver \etal model}

\begin{table}
\caption{Bubble properties at $t = 15 000 \yr$ as predicted by the
Weaver \etal (1977) model.}
\label{tab:weaver}
\begin{tabular}{ll}
Parameter                         & Value  \\  Reverse shock radius
$R_{1}$      & $ 4.2 \times 10^{18} \cm $ \\  Forward shock radius
$R_{2}$      & $ 1.1 \times 10^{19} \cm $ \\  Thermal energy in
shocked wind    & $ 1.32 \times 10^{49} \erg $\\  Kinetic energy in
shell           & $ 5.67 \times 10^{48} \erg $ \\  Shell velocity & $
143 \kmps $ \\  Total volume (including shell)    & $ 5.6 \times
10^{57} \cm^{3} $  \\  Total mass (including shell)      & $ 28.7
\Msol $ \\  Luminosity$^{a}$ (0.1-2.4 $\keV$) & $ 2.39 \times 10^{35}
\ergps $  \\  Luminosity$^{b}$ (0.1-2.4 $\keV$) & $ 8.94 \times
10^{32} \ergps $  \\
\end{tabular}
\medskip
\\ $^{a}$ Including thermal conduction. \\ $^{b}$ Without conduction.
\end{table}

The properties of our simulated bubble at $t = 15000 \yr$ and the
predictions of the Weaver \etal (1977) for the same input parameters
are encouragingly similar, if we modify the standard Weaver \etal
model to account for the {\em lack} of thermal conduction in the {\sc
VH-1} code (Fig.~\ref{fig:weaver} and Tables \ref{tab:bub_properties}
and \ref{tab:weaver}).

To illustrate the effect that conduction would have on the structure of
the bubble, the properties of the standard Weaver \etal bubble with
conduction (with thermal conductivity of the form $K(T) = CT^{5/2}$, where
$C = 1.2 \times 10^{-6} \ergps \cm^{-1} \K^{-7/2}$ (Spitzer 1962)) 
are also given in Fig.~\ref{fig:weaver} and Table~\ref{tab:weaver}. 

Note that the analytical predictions only apply for $R \geq R_{1}$,
the radius of the reverse shock, although they are shown in
Fig.~\ref{fig:weaver} extending all the way to the center of the
bubble. In Castor \etal (1975) and Weaver (1977) the free-wind region
inside $R_{1}$ is ignored, which as we shall discuss below, does lead
to minor differences between simulation and theory.

Due to the way we have injected mass and energy onto the computational
grid, the region inside $R_{1}$ corresponds to the solution of
Chevalier and Clegg (1985), rather than a point-like source of mass
and energy.

The Weaver \etal model's shell is much thinner than the {\em azimuthally
averaged} profile from the simulation, as the shell is corrugated and
hence spread out over radius. Inspecting the 2-D data the peak
densities in the simulations agree well with the density predicted by
the analytical model, implying that the dense
shell is not significantly unresolved. 
The bubble in the simulation is slightly larger
than predicted, but this is not surprising as this reflects the
``head-start'' we have given the simulated bubble by injecting mass
and energy over a finite input region.

Densities (temperatures) in the shocked-wind rise (drop) near the
shell, compared to the analytic prediction, due to the mixing of
cooler material into the bubble as discussed in
Section.~\ref{sec:results_instabs}. This increases the bubble X-ray
luminosity from $L_{\rm X} = 8.94 \times 10^{32} \ergps$ in
Table.~\ref{tab:weaver} for a non-conductive bubble to an appreciable
$6.55 \times 10^{34} \ergps$, only a factor $\sim3$ less than for a
bubble with thermal conduction.

The thermal pressure in the shocked wind in our simulation is less
than that predicted by the Weaver \etal model. In the analytic 
solution it is assumed
that the energy contained in the free-wind region is negligible, and
that all the energy in the shocked wind is thermal. In practice this
is not the case, although the majority of the energy is thermal energy
in the shocked wind. The velocities in the shocked wind are
appreciable (see Fig.\ref{fig:weaver}). Coupled with the appreciable
energy loss that accompanied shell collapse this leads to lower than
expected thermal energy in the shocked wind, and hence lower pressure.

\subsubsection{Simulated {\it ROSAT} spectra}
\label{sec:res_rosat_spectra}

We generate and fit theoretical X-ray spectra by the following
process. The 2-Dimensional density and temperature data from {\sc
VH-1}  are rotated around the axis of symmetry to produce a temporary
3-D dataset. The {\sc MEKAL} (Mewe, Kaastra \& Liedahl 1995)  plasma
code is used to calculate the X-ray emission from each volume of gas,
assuming ionisation equilibrium and solar element abundances. It is
also assumed the gas is optically thin to its own radiation, and that
any absorption the X-rays suffer is extrinsic to the source. External
absorption is modeled as a uniform foreground screen, using the
coefficients given in Morrison \& McCammon (1983).

\begin{figure}
\vspace{9.0cm} 
\includegraphics{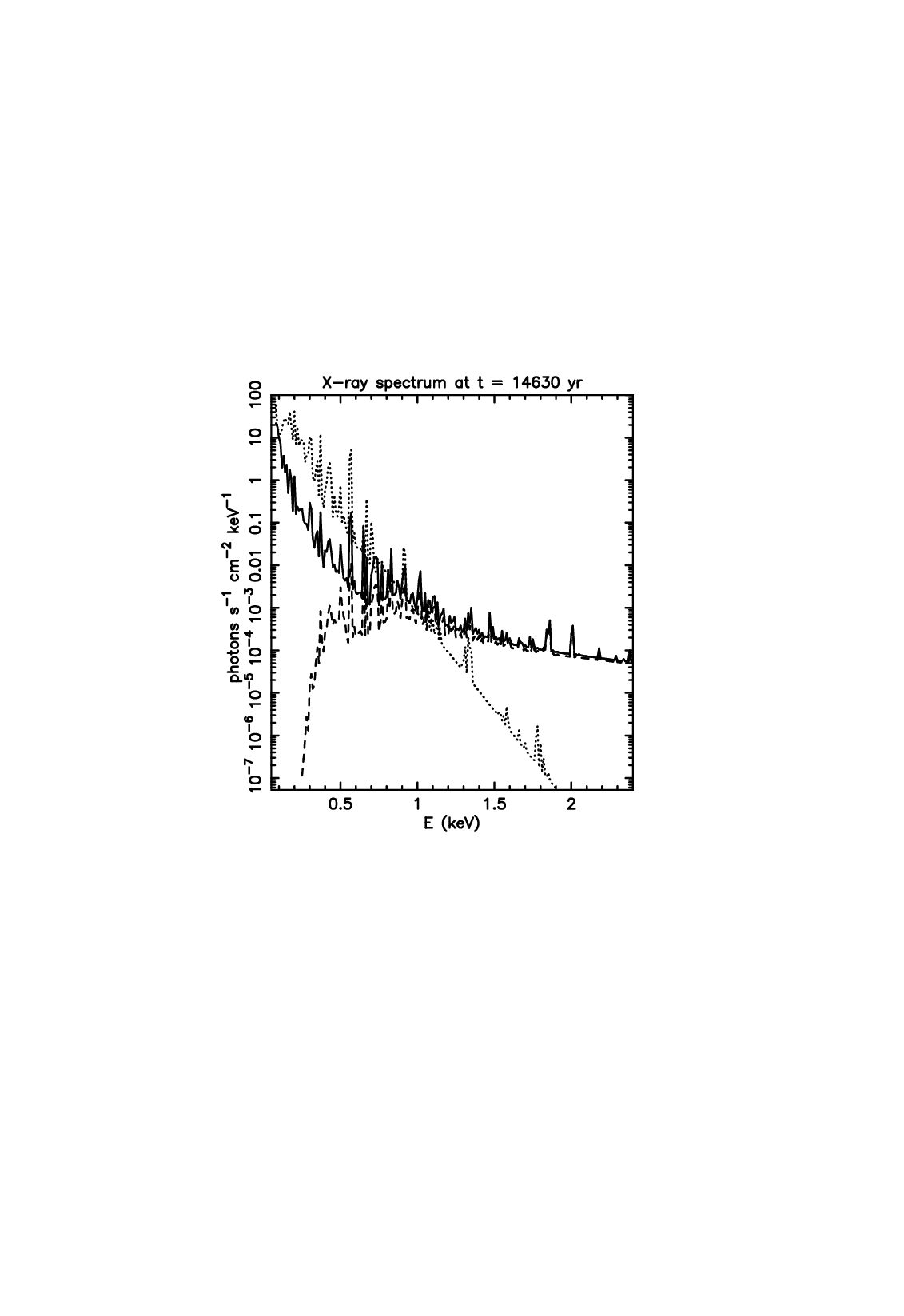}
\caption{X-ray spectra for the bubble at $t=15000 \yr$, unconvolved
with any instrument response. The solid and dashed lines correspond to
absorbing columns of $\nH = 0.0$ and $3.16 \times 10^{21} \pcm2$
respectively. For comparison an unabsorbed  
$T = 0.1 \keV$ spectrum of arbitrary
normalisation is shown dotted. Note that despite being soft the 
bubble's spectrum does have a hard tail.}
\label{fig:spectra}
\end{figure}

The raw high resolution X-ray spectrum generated by the plasma code 
(see Fig.~\ref{fig:spectra})
is corrected for a typical absorbing column of $3.16 \times 10^{21}
\pcm2$, appropriate for the wind-blown bubble NGC 6888 (Wrigge,
Wendker \& Wisotski 1994), and then folded through the spectral
response of the {\it ROSAT} PSPC, assuming a source distance of $2
\kpc$ for the bubble. 

No X-ray background is included, effectively assuming a perfect
background subtraction. We estimate that the soft X-ray background
would have a PSPC count rate of $\approx 0.19$ counts $\ps$
over the area of sky occupied by the bubble at $t = 15000 \yr$,
based on the {\it ROSAT} PSPC observation of M82 (Strickland \etal 1997). 
The detector count rate due to the bubble itself is $0.32$ counts $\ps$
for an absorbing column of $3.16 \times 10^{21} \pcm2$.

%

Note that the time to reach ionisation equilibrium, $t_{\rm ieq} \approx
10^{12} n_{\rm e}^{-1} \s$ (\cf Masai 1994) is greater than the bubble's
age over much of the volume. As our aim is to see whether fits to
{\it ROSAT} spectra can accurately reflect the true bubble properties, the
assumption of ionisation equlibrium is not a problem.

\subsubsection{A simple estimator of the fitted temperature.}
\label{sec:expected_results}
Are there easier ways of  predicting the observable properties of a model
than generating and fitting artificial X-ray spectra? This method
is also instrument specific, as it depends on the spectral characteristics
of the instrument used, and so predicted results for {\it ROSAT} are
not directly transferable to, \eg {\it ASCA}.

If we {\em assume} we can correctly estimate the absorbing column and 
metallicity, either from a spectral fit or from observations at other 
wavelengths, then estimating a characteristic temperature is all we
need do. 

We might expect the fitted temperature
to be some flux-weighted average of the true temperature range, and
the emission measure roughly the total emission over the range of
temperatures {\it ROSAT} is sensitive to.

Over the energy range $0.005$-$15.0 \keV$ the intrinsic flux (\ie
unabsorbed) weighted average temperature $<T_{EW}> =
(4.9^{+2.7}_{-1.3}) \times 10^{-3} \keV$, \ie UV emission from very
cool material at the shell-bubble interior interface. In practice all
the mid and extreme UV emission is absorbed in the intervening ISM. In
addition the instrument response further limits the ``observed''
temperature.

If we assume that we can accurately remove the effects of
interstellar absorption on the X-rays (\ie fitting the column
correctly) then to construct a predicted {\it ROSAT} temperature we
need only weight each temperature by the count rate in the {\it ROSAT}
PSPC due to each temperature component, ignoring absorption. This
temperature is $<T_{\it ROSAT}> = 0.0988_{-0.046}^{+0.268} \keV$.
Given the true temperature distribution in Fig.~\ref{fig:em} the {\it
ROSAT} emission-weighted average temperature is more a reflection of
the instrument response than a fair estimation of the temperature of
the emitted radiation or the general state of the bubble.

 $<T_{\it ROSAT}>$ is instrument specific, defeating our object of trying to
produce an instrument-independent estimator for the temperature. This may
well be impossible, in such a situation as this where the temperature 
distribution can not be described by a single, characteristic temperature.

The emission measure should follow from the normalisation of the
spectrum. and  will depend on the best-fit values for the other parameters
which determine the shape of the model spectrum. For the purposes of 
comparison between the results of the spectral fitting below and the ``true''
values, the total emission measure between $0.01 \leq T
(\keV) \leq 10.0$ is $3.77 \times 10^{58}
\pcc$. Between $0.1 \leq T (\keV) \leq 2.4$, it is $8.64 \times
10^{56} \pcc$.

\subsubsection{Spectral fitting}
\label{sec:res_spectral_fit}

For an assumed exposure time of $3000 \s$ we obtain a simulated
{\it ROSAT} PSPC spectrum containing
$\sim 1000$ counts. To assess the effect of real photon statistics,
and the consequent variation in best-fit parameters, on our simulated
spectra we generated ten Poisson realizations, fitting each using the
{\sc Starlink} X-ray analysis package {\mbox {\sc Asterix}}.  The
number of counts per bin is sufficient to allow the use of $\chi^{2}$
fitting.

Three different spectral models were considered: a standard single
temperature absorbed hot plasma model; a two temperature model, with
both components having the same absorbing column and metal abundance,
and finally a differential emission measure model. 

Single temperature models are widely used to characterise X-ray
emission, and it is only when there are sufficient counts to show that
a single temperature model is a bad fit that more complex models are
used. It is therefor sensible to fit such a model, despite knowing
that the true temperature distribution of the X-ray emitting plasma is
far from being single temperature (see Fig.~\ref{fig:em}).

A two temperature model would be the next level of complexity, naively
a soft component for cooler denser gas near the shell and a hot
component for the bubble interior. 

The differential emission model attempts to incorporate the emission
of gas at a wider range of temperatures by specifying the emission
measure $EM = \int n_{\rm e}^{2} dV$ is a power law in temperature, \ie $EM
\propto T^{\gamma}$, between two temperatures $T_{\rm low}$ and
$T_{\rm high}$. 

We quote results for models with all parameters fitted for (including
the column and the metallicity) and for fits with the metallicity
fixed at solar abundance. Although in practice we know the absorbing
column applied, fits with it fixed were statistically
unacceptable. For the purpose of display and interpretation, the
best-fit results for each of the ten Poisson realisations were
averaged. The quoted errors are the statistical deviations of the
best-fit results, and not the fitted confidence regions. Typically,
the 68\% confidence regions (for one parameter of interest) were
smaller or of order the deviations quoted.

The results for the fits to the integrated bubble spectrum are given
in Tables.~\ref{tab:rz_fits}, \ref{tab:2rz_fits} and
\ref{tab:mdm_fits}.

\begin{figure}
\vspace{9.0cm} 
\includegraphics{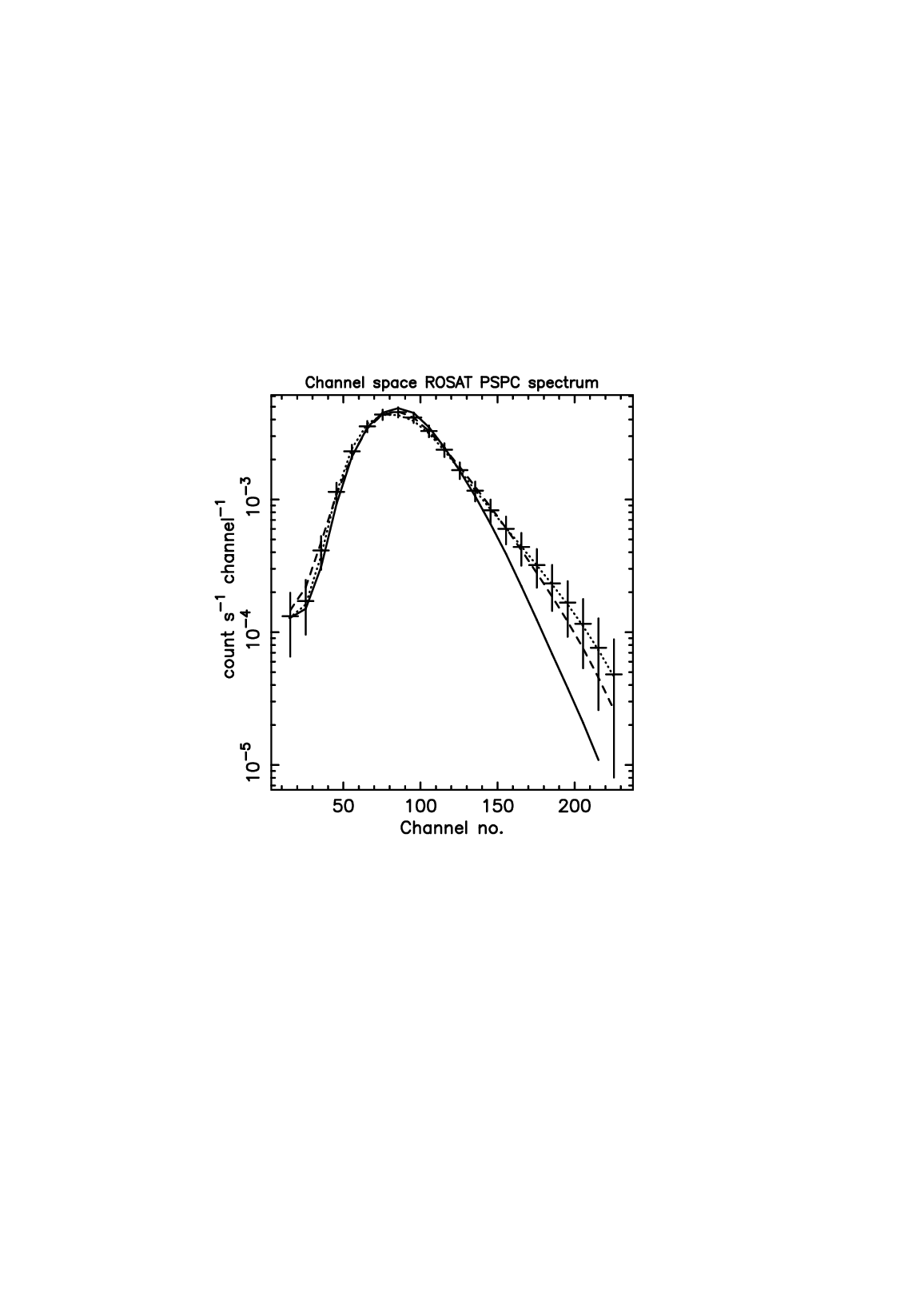}
\caption{Channel space {\it ROSAT} PSPC spectrum of the bubble at $t =
15000 \yr$ shown with the three average best fit models (single
temperature, abundance fixed at solar -- solid line; single temperature,
abundances fitted for -- dashed; two temperature, abundances fixed -- dotted)
The two temperature and differential emission measure models give very
similar {\it ROSAT} spectra.
The spectrum has no noise, the error bars represent the size of 
typical Poisson uncertainties for an exposure time of $3000 \s$.}
\label{fig:best_fits}
\end{figure}

The best fit models look very similar when convolved with PSPC
instrument response (\ie as they are when compared with the data),
despite the very different fit parameters (Fig.~\ref{fig:best_fits}). 
The only noticeable
differences are the width of the peak at channels equivalent to $E
\sim 0.8 \keV$, and how the model fits to points between $1.5 \ltsimm
E (\keV) \ltsimm 2.0$.

The reason the single temperature, metallicity fixed at solar,
spectral fits are bad is that they fail to fit these higher energy
points at all. Freeing the metallicity to fit allows the shape of the
spectrum to be changed. the temperature increases to better fit the
higher energy points, and the metallicity drops to change the relative
shape of the spectrum around the Iron-L complex (effectively the only
lines {\it ROSAT} spectral fitting is sensitive too). Similarly the
absorption column changes the shape of the spectrum. In the fits with
metallicity free to fit, the increased temperature move the spectrum
towards higher energies,  removing the requirement to absorb as much
low energy flux, hence low best fit columns.

The two temperature and differential emission measure models give better
fits, as the hot component (or hot components in the case of the
differential emission measure model) fit the high energy points
present in the spectrum, freeing the cool component to fit at a lower
temperature. The available trade-off between the various components in
these multi-component models make the best-fit parameter less
constrained, and insensitive to the metal abundance.

We shall defer further interpretation and discussion of these results,
and what they would lead us to infer about the bubble's properties, to
Section.~\ref{sec:disc_obs_imp}.

\begin{table}
\caption{Average result of fits to simulated {\it ROSAT} spectra using
a single  temperature {\sc MEKAL} hot plasma model with
absorption. The errors quoted are the deviations between the best fit
results. See text for details.  Note the significant improvement in
fit when metal abundance is free to fit.}  
\label{tab:rz_fits}
\begin{tabular}{lll} 

Parameter & Single temperature fits & Units \\ 

\hline

$\nH$          & $ 7.8 \pm{ 0.4 }$     & $10^{21} \pcm2$ \\ 
$EM$           & $ 0.018 \pm{ 0.005 }$ & $10^{60} \pcc$ \\   
$T$            & $ 0.163 \pm{ 0.006 }$ & $\keV$ \\   
$Z$            & $ 1.0 $ (Fixed)       & $Z_{\odot}$ \\   
$f_{X}$ (obs.) & $ 2.86\pm{0.07}$      & $10^{-12} \ergps \pcm2$ \\   
$f_{X}$ (int.) & $ 7.18\pm{1.84}$      & $10^{-10} \ergps \pcm2$ \\   
$\chi^{2}$ (d.o.f) & $ 39 \pm{ 5.6 }$ ( 19 )     & \\  

\hline

$\nH$         & $ 1.9 \pm{ 0.4 }$       & $10^{21} \pcm2$ \\ 
$EM$          & $ 0.0019 \pm{ 0.0012 }$ & $10^{60} \pcc$ \\   
$T$           & $ 0.45 \pm{ 0.07 }$     & $\keV$ \\   
$Z$           & $ 0.04 \pm{ 0.02 }$     & $Z_{\odot}$ \\ 
$f_{X}$ (obs.)     & $ 3.29\pm{0.10}$   & $10^{-12} \ergps \pcm2$ \\   
$f_{X}$ (int.)     & $ 0.17\pm{0.07}$   & $10^{-10} \ergps \pcm2$ \\   
$\chi^{2}$ (d.o.f) & $ 18.5 \pm{ 4.8 }$ (     18 ) & \\ 

\hline

\end{tabular}
\end{table}

\begin{table}
\caption{As Table~\ref{tab:rz_fits} but fitting with a two temperature
{\sc MEKAL} model.}  
\label{tab:2rz_fits}

\begin{tabular}{lll}

Parameter & Two temperature fits & Units \\ 

\hline

$\nH$        & $ 3.5 \pm{ 2.8 }$         & $10^{21} \pcm2$ \\
$EM_{1}$     & $ 0.04 \pm{ 0.10 }$       & $10^{60} \pcc$ \\   
$T_{1}$       & $ 0.23 \pm{ 0.08 }$       & $\keV$ \\   
$EM_{2}$     & $ 0.00013 \pm{ 0.00006 }$ & $10^{60} \pcc$ \\ 
$T_{2}$      & $ 1.28 \pm{ 0.30 }$       & $\keV$ \\   
$Z$          & $ 1.0 $ (Fixed)           & $Z_{\odot}$ \\
$f_{X}$ (obs.)     & $ 3.32\pm{0.10}$    & $10^{-12} \ergps \pcm2$ \\   
$f_{X}$ (int.)     & $ 13.7\pm{32.9}$    & $10^{-10} \ergps \pcm2$ \\   
$\chi^{2}$ (d.o.f) & $ 12.0 \pm{ 4.6 }$ ( 17 )   & \\ 

\hline

$\nH$        & $ 3.7 \pm{ 2.9 }$         & $10^{21} \pcm2$ \\ 
$EM_{1}$     & $ 0.013 \pm{ 0.031 }$     & $10^{60} \pcc$ \\   
$T_{1}$      & $ 0.24 \pm{ 0.11 }$       & $\keV$ \\   
$EM_{2}$     & $ 0.00005 \pm{ 0.00006 }$ & $10^{60} \pcc$ \\  
$T_{2}$      & $ 1.45 \pm{ 0.37 }$       & $\keV$ \\   
$Z$          & $ 6.7 \pm{ 5.0 }$         & $Z_{\odot}$ \\   
$f_{X}$ (obs.)     & $ 3.31\pm{0.11}$    & $10^{-12} \ergps \pcm2$ \\  
$f_{X}$ (int.)     & $ 33.3\pm{80.9}$    & $10^{-10} \ergps \pcm2$ \\   
$\chi^{2}$ (d.o.f) & $ 11.5 \pm{ 4.1 }$ ( 16 )    & \\  

\hline

\end{tabular}
\end{table}


\begin{table}
\caption{As Table~\ref{tab:rz_fits} but fitting with a differential
 emission measure model, where the emission measure at any temperature
 $EM_{\rm (T)} \propto T^{\gamma}$, between two cut-off temperatures
 $T_{\rm low}$ and $T_{\rm high}$. This model attempts to recreate 
the multi-phase origin of the emission.}  
\label{tab:mdm_fits}

\begin{tabular}{lll}

Parameter & Diff. emission measure fits & Units \\

\hline

$\nH$         & $    3.6 \pm{    2.3 }$           & $ 10^{21} \pcm2 $ \\ 
$EM$          & $ 0.00376 \pm{ 0.00524 }$         & $10^{60} \pcc$ \\   
$\gamma$      & $ -0.23 \pm{   0.36 }$            & \\   
$T_{\rm low} $    & $  0.0118 \pm{  0.0345 }$     & $\keV$ \\  
$T_{\rm high}$    & $    7.4 \pm{    2.2 }$       & $\keV$ \\   
$Z$               & $ 1.0 $ (Fixed)               & $Z_{\odot}$ \\ 
$f_{X}$ (obs.)     & $ 3.37\pm{0.13}$      & $10^{-12} \ergps \pcm2$  \\   
$f_{X}$ (int.)     & $ 1.46\pm{2.03}$      & $10^{-10} \ergps \pcm2$ \\   
$\chi^{2}$ (d.o.f) & $    15.2 \pm{ 6.3 }$ (     17 ) & \\ 

\hline

$\nH$          & $    3.0 \pm{    1.8 }$      & $10^{21} \pcm2$ \\ 
$EM$           & $ 0.00264 \pm{ 0.00287 }$    & $10^{60} \pcc$ \\   
$\gamma$       & $ -0.29 \pm{   0.39 }$       &  \\  
$T_{\rm low}$     & $ 0.0071 \pm{  0.0193 }$  & $\keV$ \\  
$T_{\rm high}$    & $    8.7 \pm{    4.6 }$   & $\keV$ \\  
$Z$               & $ 1.16 \pm{ 1.53 }$       & $Z_{\odot}$ \\  
$f_{X}$ (obs.)     & $ 3.38\pm{0.12}$         & $10^{-12} \ergps \pcm2$ \\
$f_{X}$ (int.)     & $ 0.81\pm{1.33}$         & $10^{-10} \ergps \pcm2$ \\  
$\chi^{2}$ (d.o.f) & $ 14.2 \pm{ 5.7 } $ (     16 ) & \\  

\hline

\end{tabular}
\end{table}

\begin{figure}
\vspace{8.0cm} 
\includegraphics{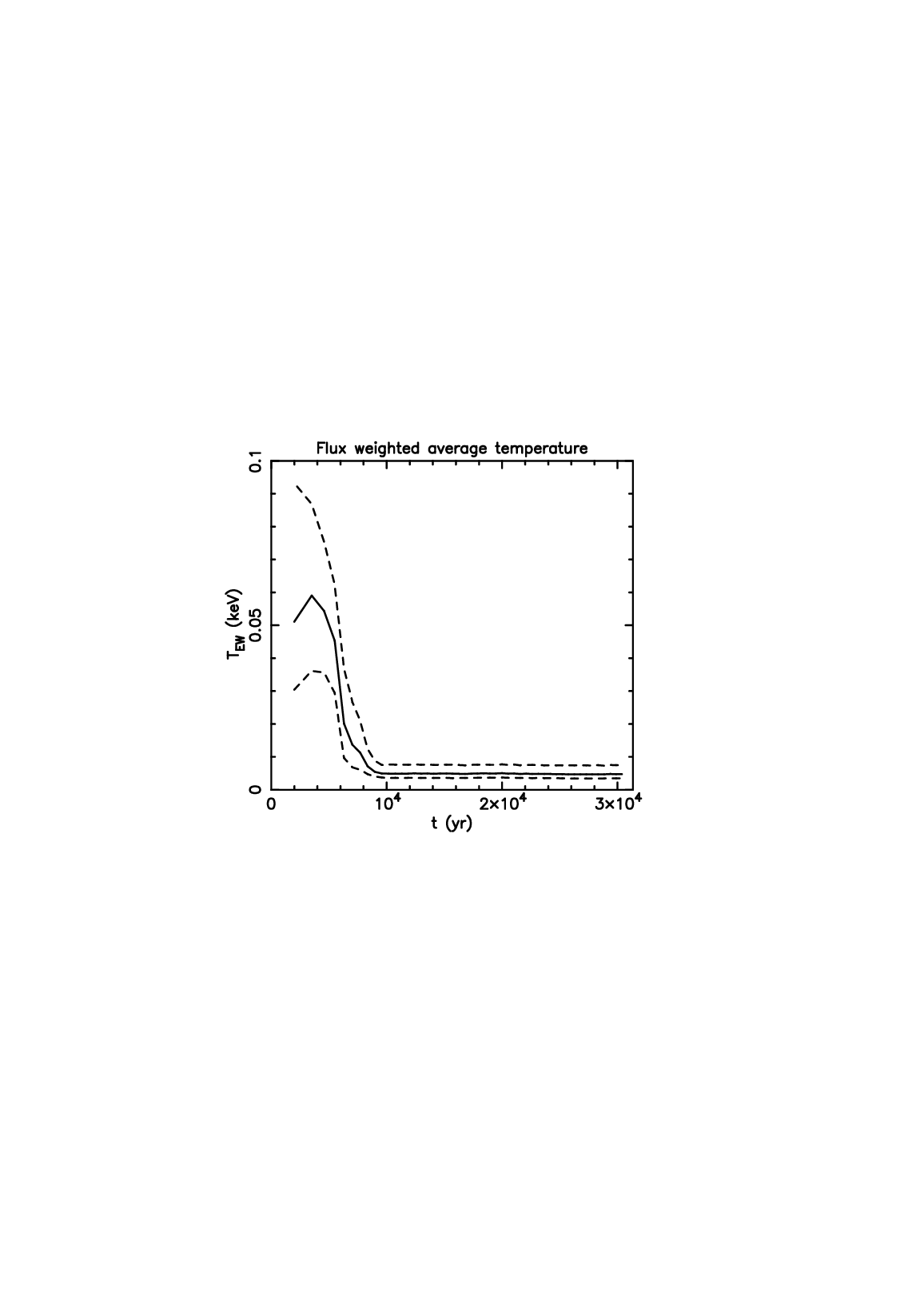}
\caption{The flux weighted average temperature $T_{\rm EW}$ of the
bubble (see Sect.~\ref{sec:disc_spec}), as a function of the age of
the bubble, in the band $0.005$-$15.0 \keV$. At early times the
emission is predominantly from the swept-up and shock heated
ISM. After this shell cools the total bubble emission is 
still dominated by the
shell, but the radiation is now UV rather than soft X-rays.
Dashed lines show the r.m.s. deviation of $T_{\rm EW}$.}
\label{fig:t_ew}
\end{figure}

\begin{figure}
\vspace{9.0cm} 
\includegraphics{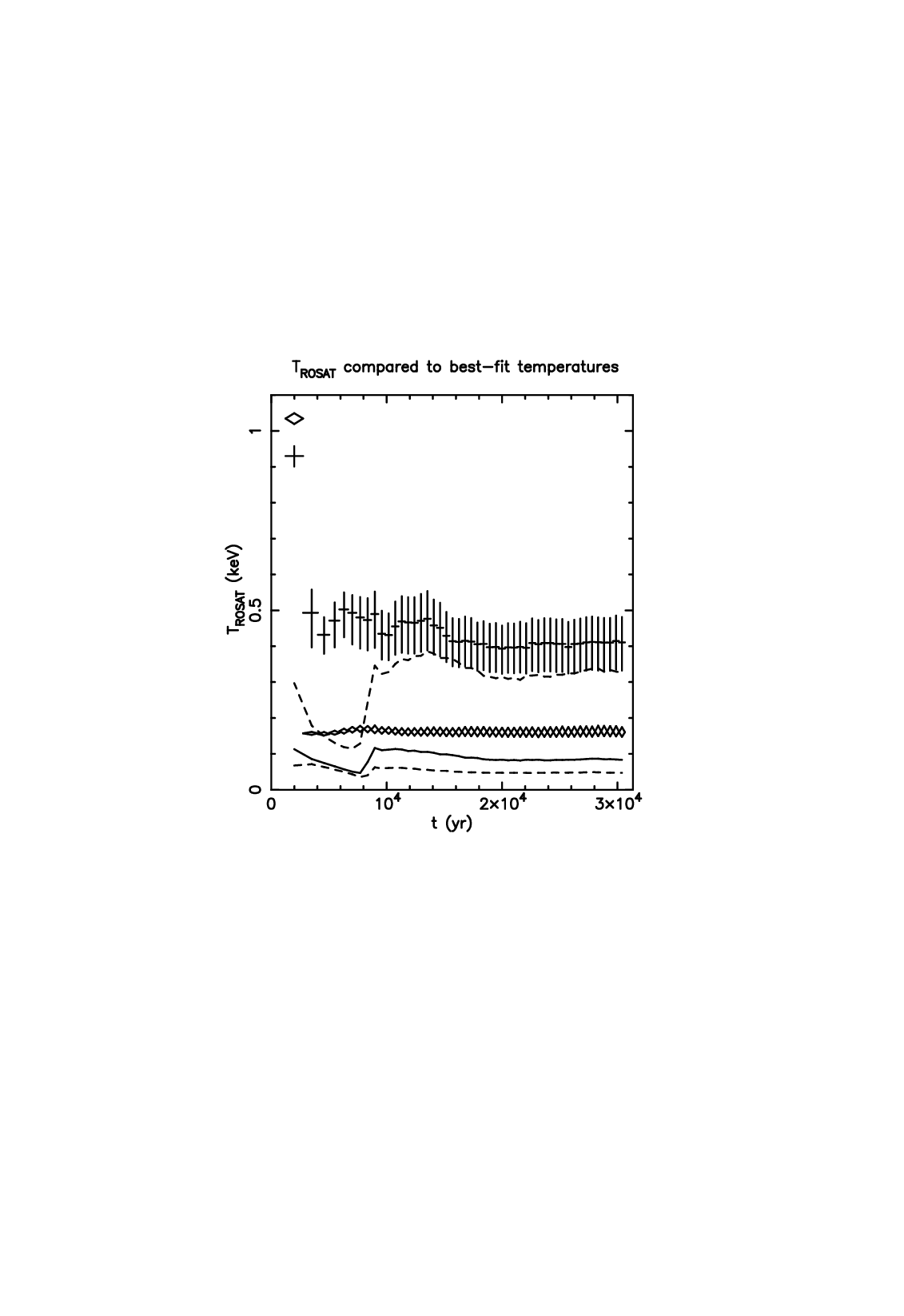}
\caption{The {\it ROSAT} flux-weighted average temperature $T_{\rm
ROSAT}$ as a function of bubble age (see Sect.~\ref{sec:expected_results}),
compared the temperatures  obtained from single temperature fits to
the PSPC spectra (Diamonds:  fits with metal abundance fixed at solar,
crosses: fits with metal  abundance free to fit). Error bars are
68\% confidence in one parameter of interest. The dashed lines show
the r.m.s. deviation of $T_{\rm ROSAT}$.}
\label{fig:t_rosat}
\end{figure}

\subsection{Spectral variation over time}
The spectral properties of the bubble after shell collapse will be
different to those prior to shell cooling. At early times, as
discussed above, the shell is the major source of X-rays, and
dominates the emission. It is only after the shell cools from its
initial temperature of $T \sim 10^{6} \K$ to the final temperature of
$T \sim 10^{4} \K$ that the X-ray emission from the bubble interior,
\ie the shocked wind itself, becomes significant.

This is graphically illustrated by Figs.~\ref{fig:t_ew} and
\ref{fig:t_rosat}. Note how after shell collapse the averaged
temperature contributing to the photons detected by {\it ROSAT},
$T_{\rm ROSAT}$, increases again. This is the emission from shocked
wind itself, previously of too low a level compared to the X-ray
emission from the shell to be noticeable.

Fig.~\ref{fig:t_rosat} shows the best fit temperatures from single
temperature fits to a set of simulated {\it ROSAT} spectra over time,
for both metal abundance fixed at solar and abundances free to fit. As
generating 10 Poisson realisations for each of $\sim 50$ spectra, fitting 
and finally averaging the results would be prohibitively
time consuming, we have fitted the spectra with no Poisson noise
applied.  Comparison of this with the more rigorous method used
earlier shows that this generates essentially identical results,
although the values of $\chi^{2}$ obtained are misleadingly low.

As can be seen from Fig.~\ref{fig:t_rosat} the best fit temperatures 
fail to reflect the
temperature changes we know are occurring. Intriguingly, the best fit
metallicity does show a systematic trend between $t \sim 5000$ and $10
000 \yr$. Two temperature fits, able to fit the shape of the spectrum
better, do show an initial drop in both temperatures to a minimum at
$t \sim 5000 \yr$, followed by a increase to a constant value after $t
\sim 9000 \yr$ (although the hotter component only levels out after $t
\sim 15 000 \yr$). 

\subsection{The high mass loss rate simulation}
\label{sec:res_cool_bub}
To further investigate how spectral fitting to the observed X-ray data
depends on the the properties of the bubble, we have repeated the detailed
analysis described in Section~\ref{sec:res_t15000} 
on the higher wind mass loss rate simulation.

Increasing $\Mdot_{\rm W}$ to
$10^{-4} \Msol \pyr$ results in a bubble with the same size at 
$t = 15 000 \yr$, identical cold shell properties, but a shocked wind
with a density (temperature) a factor $2$ higher (lower).

Single temperature models give best-fit results very similar to those
given in Table~\ref{tab:rz_fits} for the low mass loss rate simulation,
except the best fit temperatures are significantly higher in this case:
$T = 0.41\pm{0.09} \keV$ (metal abundance fixed) or 
$T = 0.60\pm{0.13} \keV$ (metal abundance fitted for). This is despite
the average temperature within the shocked wind being half that of the
low mass loss rate simulation. Again absorption columns (and when fitted 
for, metallicities) deviate systematically from the true values.

The two temperature spectral models give best fits very similar to the
results given in Table~\ref{tab:2rz_fits}, the best fit temperatures being
$T_{1} = 0.26\pm{0.07} \keV$ and $T_{2} = 1.47\pm{0.31} \keV$. The only
real differences are a) the relative fraction of the total emission 
measure in the hot component has increased (\ie the spectrum appears to be
harder), and b) there is evidence for two fit minima, one with low $\nH$ and
emission measure, the other a high $\nH$ with high emission measure.

\section{Discussion}
\label{sec:disc}

\subsection{Properties inferred from the spectral fitting}
\label{sec:disc_spec}
We shall concentrate on the results from the low mass loss rate simulation,
mentioning only where the spectral fit results from the high mass loss
rate simulation differ.

\subsubsection{Single temperature models}
Single temperature spectral fits give misleading information on the
state of the bubble, not surprising given the multicomponent
temperature structure and the detector characteristics, nor are they
sensitive to the true spectral changes that occur during the bubble's
growth (see Fig.~\ref{fig:t_rosat}). This suggests that the best-fit
values are strongly affected by the multicomponent structure and the
instrument response.

The confidence levels from the fits misrepresent the true
uncertainties in the fitted parameters. In particular the fitted
temperatures are apparently well constrained, although the emission
comes from a very broad temperature distribution. The spectral fits give
little clue to the true temperature distribution, as the
fits with metallicity free to fit have acceptable $\chi^{2}$. A naive
hope that in a  situation such as this, with a broad, non-peaked
temperature distribution, single temperature fits would show poorly
constrained best-fit temperatures is not justified by these results.

Absorbing columns and metal abundances are both systematically
mis-fitted. It is clear that by fitting a multicomponent spectrum with
a single temperature model we introduce a major systematic
effect. Single temperature fits with metal abundance fixed at solar
(Table~\ref{tab:rz_fits}) are generally unacceptable, with reduced
$\chi^{2} \sim 2$, perhaps indicating that the model is not a good
representation of the data. The hydrogen columns deviate significantly
from the true absorbing column. 

Freeing the metal abundance to fit
does give good fits to the data, but at the expense of giving best-fit
parameters that bear little relation to the bubble's true
properties. The best fit metal abundances are less than one twentieth
solar, and apparently strongly constrained.
This is interesting given the current debate over the accuracy of
X-ray determined abundances (\cf Bauer \& Bregman 1996).

The flux-weighted average temperature $T_{\it ROSAT}$ derived in 
Section~\ref{sec:expected_results} is not a good estimator of the
best-fit temperature, as can be seen in Fig.~\ref{fig:t_rosat}.
An average temperature, even sensibly weighted, does not reflect how
spectral fitting works, and will not give acceptable results. 

\subsubsection{Two temperature models}
These (Table~\ref{tab:2rz_fits}) give significantly better fits to the
data than the single temperature models and are slightly better than the
differential emission measure models. The absorbing columns fit closer
to the true value. The fits are insensitive to the metallicity, and
would not force us to believe they are significantly different from
solar, unlike the single temperature models.

That the best-fit temperatures in the two temperature model fits to the
low and the high mass loss rate simulations are so similar, with only the
relative normalisation of the two components varying, suggests the
temperatures are determined more by the {\it ROSAT} PSPC's response than
the true temperature distribution of the source.

\subsubsection{Differential emission measure models}
From Fig.~\ref{fig:em} it is clear that for $T \ltsimm 0.7 \keV$ the
emission measure is well represented by a power law in $T$, with a
slope of $\gamma \approx -1.8$. Between $T = 0.1 \keV$ and $T = 0.7
\keV$ the true emission measure falls by nearly two orders of magnitude, so
a differential emission measure model with this slope should provide a good
approximation to the spectrum.

The results of the model fits (Table~\ref{tab:mdm_fits}) are therefor
surprising, given the best fit values for $\gamma \approx -0.25$! The
combination of material hotter than $T \sim 1 \keV$ and the energy
dependent response of the {\it ROSAT} PSPC must bias the fit
significantly. The slope is clearly wrong within the energy range
$0.1$-$2.4 \keV$ ROSAT is sensitive to, let alone extended to the
cut-off temperatures claimed by the fit.

\subsubsection{The effect of longer exposure times}
Are any of the effects above due to low numbers of photons? Repeating
the above analysis with assumed exposure times of $10000 \s$, giving
simulated spectra with $\gtsimm 3000$ counts, gives results very similar
to those in Section~\ref{sec:results}. The confidence regions quoted on
 the individual fits are smaller, but the average best fit result is 
consistent with those quoted above, and are statistically acceptable fits. 

It therefor appears that the systematic deviations of the spectral fit results
from the true values is due to fitting an intrinsically complex spectrum with
a simplistic model, and not due to poor photon statistics.

\subsubsection{Inferred properties}
\label{sec:disc_obs_imp}

Can we infer any of the true bubble properties from the spectral fits?
In particular, the true luminosity, densities, pressures and thermal
energy are interesting quantities that we would like to know in
addition to the temperature if \mbox{X-ray} observations are to be of any use 
in understanding the object.

The $0.1$-$2.4 \keV$ fluxes predicted from all the models are
reasonably accurate {\em before correction for absorption} (compare the real
values from Table ~\ref{tab:bub_properties} with the results from the
spectral fitting, Tables~\ref{tab:rz_fits} - \ref{tab:mdm_fits}). The
single temperature models get the intrinsic luminosity very wrong, due
to the incorrect absorption column, either overestimating it by a
factor $\sim5$ (metallicity fixed at solar) or underestimating by an
order of magnitude in the case of the fits with the metallicity free
to fit. On average the two temperature models overestimate the
intrinsic flux by an order of magnitude, although with a large
uncertainty due to the large variation in best-fit column. Similarly
the differential emission measure models also have a large scatter in
inferred intrinsic luminosity.

Assuming we know the volume of the emitting plasma $V$, the root mean
square electron density is $n_{\rm e} = \sqrt(EM/V)$, where $EM$ is the
(volume) emission measure emission measure obtained from the spectral fit. 
Then the thermal pressure and energy are $P
\approx 2 n_{\rm e} k T$ and $E_{\rm TH} \approx 3 n_{\rm e} k T V$.

\begin{figure}
\vspace{8.0cm} 
\includegraphics{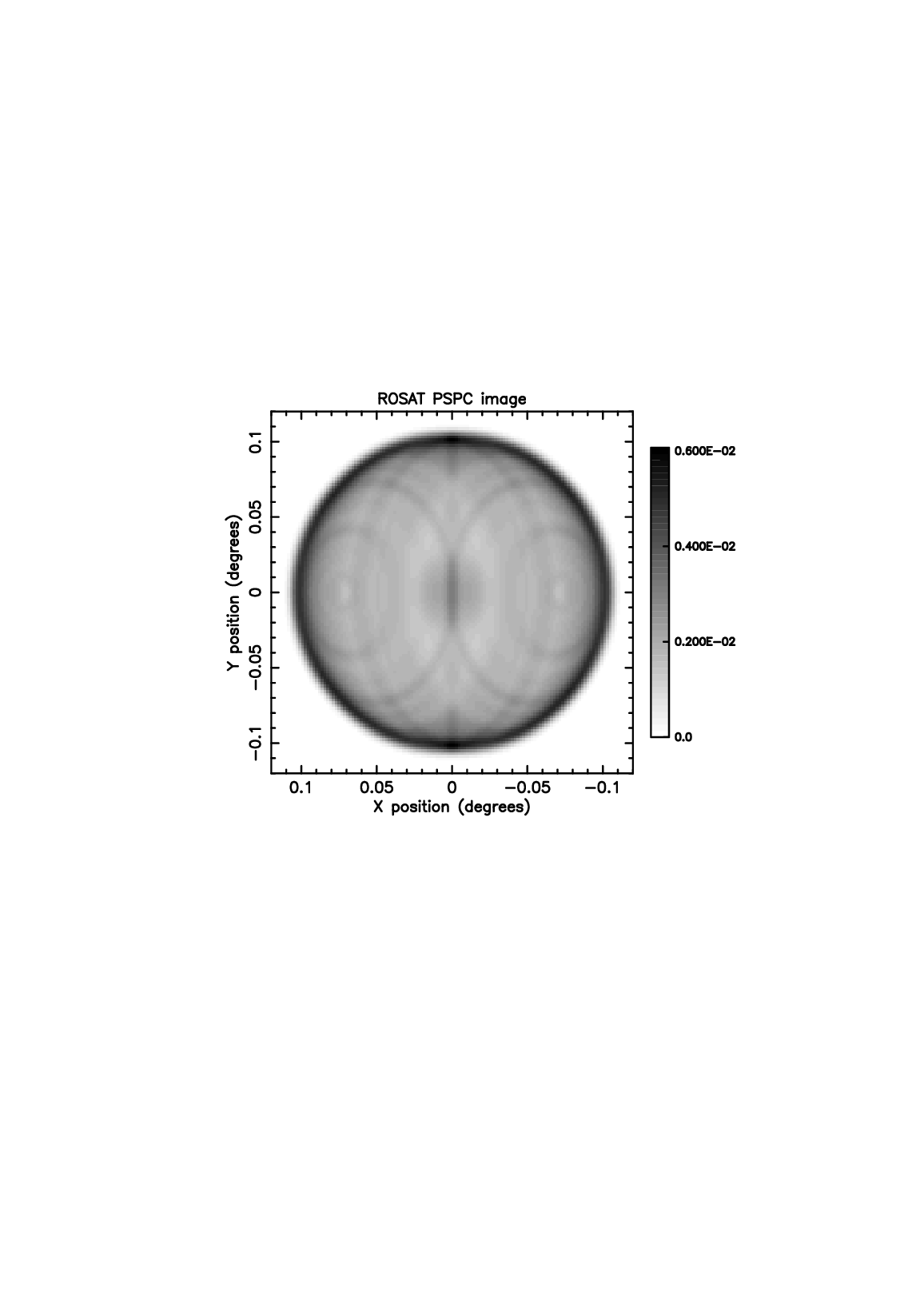}
\caption{An X-ray image of the bubble at $t=15000 \yr$, in the $0.1$-$2.4
\keV$ energy band. The image has been smoothed with a Gaussian of FWHM
$25\arcsec$ to approximate the {\it ROSAT} PSPC PSF. Units are counts
$\ps {\rm arcmin}^{-2}$.}
\label{fig:ximage}
\end{figure}

Given the X-ray surface brightness (Fig.~\ref{fig:ximage}) 
we can estimate the volume. The inferred properties are relatively
insensitive to $V$ as $P$ and $n_{\rm e}$ are proportional to
$V^{-1/2}$ and $E_{\rm TH} \propto V^{1/2}$.

The bubble is clearly limb-brightened (Fig.~\ref{fig:ximage}), the emission 
coming predominantly from a thin shell near the shocked-wind/cold shell
interface. From a radial profile of the {\it ROSAT} surface
brightness, and correcting for the point spread function, the inferred
radial thickness of this shell is $\sim 20''$,  the PSF-corrected
volume $V_{\rm corr} = 9.8 \times 10^{56} \cm^{3}$. By way of
comparison, the total volume within the bubble inferred from the X-ray
image is $V_{\rm bub} = 6.0 \times 10^{57} \cm^{3}$. Note that this is
less than the volume quoted in Table~\ref{tab:bub_properties} as we
are not including the cold shell.

The density, pressure and thermal energy inferred from the single
temperature models, using $V_{\rm corr}$ and $V_{\rm bub}$, are
given in  Table~\ref{tab:inferred}.

For the two temperature models, it would be sensible to associate the
cool component (with a high emission measure) of the spectral fit, to
cool, dense material near the shocked-wind/cold shell interface, given
the observed limb brightening. The hot component in the fit, with a
total emission measure two to three orders of magnitude less than the
cool component, could well represent emission from the hot
rarefied gas  in the bubble interior. Assigning volumes $V_{\rm corr}$
and  $V_{\rm bub}$ to the cool and hot components respectively we
obtain densities, pressures and thermal energy content given in
Table~\ref{tab:inferred}.  

From the simulation we know the true properties of all the material 
in the bubble.
The density in the plasma contributing the majority of 
the $0.1$-$2.4 \keV$ flux (all the
material with volume emissivities within an order of magnitude 
of the maximum emissivity) varies between $2.5 \ltsimm n_{\rm e} 
\ltsimm 28.2 \pcc$, averaging $7.8 \pcc$. In the hotter bubble 
interior the electron density is $n_{\rm e} \approx 0.09 \pcc$.
The shocked wind and cold shell are practically isobaric at 
$P = 1.3 \times 10^{-9}$ dyn $\pcm2$, and the majority 
of the total thermal energy in 
the bubble of $E_{\rm TH} = 1.17 \times 10^{49} \erg$ is in the hot
shocked wind material that occupies most of the bubble volume.

Comparing the inferred properties of the bubble from
Table~\ref{tab:inferred} with the true properties (see for example
Fig.~\ref{fig:weaver}) we find:
\begin{itemize}
\item The single temperature spectral 
fits, assuming the emission come from a shell of volume $V_{\rm corr}$,
give, on average,  estimates of the density and pressure (in the 
plasma that dominates the X-ray emission detected) that are 
within a factor $2$ of the true values.
\item Thermal energies inferred from single temperature models generally
underestimate the true thermal energy in the bubble. The single temperature 
models are dominated by the cool gas, occupying only a fraction of the 
total volume,  whose properties are not typical 
of the bubble as a whole, and miss the hotter gas that
contains most of the thermal energy.
\item Two temperature models give a better idea of the bubble properties,
although the temperature of the hot component is significantly lower than 
the true temperature within the shocked wind.
\item By attempting to fit a intrinsically 
multicomponent spectrum with simplistic 
single temperature model, the best fit hydrogen column and metal abundances
can be significantly in error. In more complex models this effect
is reduced, but not eradicated. For the three spectral models considered
the fitted temperatures give little indication of the true temperature
distribution.
\item Spectral fits give a good estimate of the observed energy flux, but
extrapolating the intrinsic source flux is error prone due to its
strong dependence on the assumed absorption column, and may be an order
of magnitude out. 
\item The results of the spectral fitting give little indication of the
true variation in properties within the bubble, \ie the multiple
temperature components. In particular, components with temperatures outside
the primary bandpass of the instrument are not represented by the spectral 
fitting. The differential emission measure model fails to give a better fit
than a two temperature model, although on paper it should perform better.
There would be little reason to believe that the two temperature model was
in fact an over simplification of the true situation.
\item The systematic error introduced by under-modeling the true spectrum
will dominate over any statistical error on the best-fit parameters. The
best fit may be a statistically good fit, but does not give any indication 
that it is a poor representation of the true properties.
\end{itemize}

It is interesting to note that despite the difference between the 
true bubble properties and those for a bubble with thermal conduction
(compare the radial profiles in Fig.~\ref{fig:weaver}), the properties
inferred from the {\it ROSAT} PSPC two temperature spectral fitting
are generally consistent with a bubble with conduction:

\begin{itemize}
\item The predicted luminosity of a conductive bubble is slightly higher 
than that of the simulated bubble, but the spectral fits generally 
overestimate the intrinsic luminosity of the bubble.
\item The temperatures predicted by the model with thermal conduction
are $T \sim 1 \keV$ in the shocked wind, dropping to $T \sim 0.2 \keV$ 
in the denser material near the boundary with the cold shell
(Fig.~\ref{fig:weaver}), very 
similar to the best fit temperatures.
\item The mass flux off the cold shell driven by thermal conduction
naturally leads to a region of warm X-ray emitting gas of the right
density just inside the cold shell.
\item Only the density of the hot component is inconsistent with the 
value predicted by the model with thermal conduction.The inferred 
density in the hotter material is too low by a factor $\sim 5$.
\end{itemize}

\begin{table}
\caption{Bubble properties as would be inferred from the {\it ROSAT}
spectral fits to the low mass loss rate simulation. 
The volumes refer to the volumes defined in the text.}
\label{tab:inferred}
\begin{tabular}{lllll}
Model & Volume & $ n_{\rm e}$ & $ P$ & $ E_{\rm TH} $ \\  &         &
 ($\pcc$) & (dyn $\pcm2$) & ($\erg$) \\ 

\hline 

1T, $Z=1$ & $ V_{\rm corr} $ &  $ 4.3 $ & $ 2.2 \times 10^{-9} $ & $
	3.3 \times 10^{48} $ \\ 

	& $ V_{\rm bub} $ & $ 1.7 $ & $ 9.0 \times 10^{-10} $ & $ 8.2
	\times 10^{48}$ \\

1T, $Z$ free & $ V_{\rm corr}$ & $1.4$ & $2.0 \times 10^{-9}$ & $2.9
	\times 10^{48}$ \\

	& $ V_{\rm bub}$ & $0.6$ & $8.0 \times 10^{-10}$ & $7.3 \times
	10^{48}$ \\

2T, $Z=1$ & & & & \\ 
(cool) & $V_{\rm corr}$ &
	$6.6$ & 
	$4.8 \times 10^{-9}$ & 
	$7.1 \times 10^{48}$ \\ 

(hot)  & $V_{\rm bub}$ & 
	$0.15$ &
	$6.1 \times 10^{-10}$ & 
	$5.5 \times 10^{48}$ \\

2T, $Z$ free & & & & \\ 
(cool) & $V_{\rm corr}$ & 
	$3.1$ & 
	$2.9 \times 10^{-9}$ & 
	$4.3 \times 10^{48}$ \\ 
	
(hot)  & $V_{\rm bub}$ & 
	$0.09$ & 
	$4.3 \times 10^{-10}$ & 
	$3.9 \times 10^{48} $ \\

\hline

\end{tabular}
\end{table}

\subsubsection{High mass loss rate simulation}
The true properties of the bubble in the high mass loss rate simulation
are practically identical to those of the low mass loss rate bubble. The
only differences are that the density and temperature in the shocked wind
are a factor of two different. The intrinsic X-ray luminosity in the 
$0.1$-$2.4 \keV$ band is 3\% higher, but most importantly the {\it ROSAT}
count rate is double that of the low mass rate simulation.

The intrinsic luminosity is again dominated
by the cool, dense material near that shocked wind/cold shell interface,
whose properties are almost identical between the two simulations.

The count rate depends on both the intrinsic spectrum and the
detector's spectral response.
In this case the cooler bubble interior brings the emission from the
shocked wind more within the {\it ROSAT} spectral response, giving more
high energy photons in the PSPC spectra, and hence making them seem harder.
This explains the higher best-fit temperatures of this cooler bubble,
compared to the low mass loss rate simulation.

This is further evidence, if needed, that it is necessary to predict the 
{\em observable} properties of a model when comparing observation with 
theoretical models.

\subsubsection{Summary of ``observed'' properties}

Fitted absorption columns can vary wildly in a case with a complex 
spectrum, biasing estimated emission measures are luminosities. It
might be safer not to fit for the column, and use estimates from
optical or \hi measurements.

The metallicity in single temperature models is badly underestimated.
Given the current debate over X-ray determined abundances 
(Bauer \& Bregman 1996), where
much of the attention has concentrated on the hot plasma codes rather
than the effect of multi-temperature gas, this is a very interesting
result. Is this a peculiarity of this particular model, or a general
property? Does this also affect higher spectral resolution instruments
such as {\it ASCA} or the future {\it XMM}?

As single temperature fits fail to represent to true emitted spectrum
they generally underestimate the thermal energy in the bubble, as this
is primarily in hot, tenuous gas with much lower emissivity 
than the cool gas. We note that this could provide an explanation for the
apparent discrepancy reported by Magnier \etal (1996) between
the thermal energy in the superbubble N44 as derived from {\it ROSAT} 
and {\it ASCA} spectral fits, where the X-ray derived thermal 
energy is a factor $6$-$10$ lower than expected on the basis of the
Weaver \etal model. However, we have not
demonstrated that our results apply over such a wide range
of parameter space, and the alternative 
explanation that the stellar wind energy input may have been 
overestimated may well be true. We are continuing this work to 
specifically model superbubbles in future papers.

Nevertheless, it is safe to say that the results of simple 
spectral fits to what are in reality more complex
sources depend strongly on both the spectral response of the detector
and the spectral distribution of the source.

\subsection{Limitations and assumptions}
\label{sec:disc_limits}
We have not included several physical effects that are potentially
important in real stellar wind blown bubbles: thermal conduction,
magnetic fields, time-dependent ionisation, non-solar abundance
ratios, mass loading and interaction with the winds from the main
sequence and red giant stages of the star. Mass and energy were
injected onto the computational grid in a manner more suitable for
superbubbles around OB associations and in starburst galaxies. For the
spectral fitting, we have ignored the soft X-ray background.

These omissions do not invalidate or reduce the significance of the
results contained in this paper. Our aim is to illustrate the need for
modeling the X-ray emission, and analysing in as similar a manner to
real observations as possible. In this introductory paper we have avoided
the potential complications that additional physics would introduce into
the interpretation of the results, and
take the known properties from relatively simple model, 
calculate what would be seen in a
real observatory such as {\it ROSAT}, and ask ``can we infer the true
properties from the X-ray emission resulting from those properties?''
Relaxing the above assumptions only make the need for this method of
direct comparison more urgent.

\subsection{Future work}
We intend to extend this work, relaxing the assumptions of constant
mass and energy injection, and a homogeneous ISM, to model the properties
of superbubbles and galactic winds in starburst galaxies. As a general
method and philosophy it can, of course, be applied to a wider range 
of astrophysical bubbles.

In future we shall consider the spatial variation of the X-ray properties,
as done for example for the {\it ROSAT} observation of M82's
galactic wind by Strickland, Ponman \& Stevens (1997).

We shall also extend this work to other X-ray observatories such as 
{\it ASCA} or {\it XMM}, to make use of the superior spectral resolution
of these instruments. For the higher signal-to-noise spectra that {\it XMM}
will make available, this method of direct comparison will be absolutely
necessary!

\section{Conclusions}
\label{sec:conclusions}

We have shown that in order to compare X-ray observations to theory, 
it is necessary to
consider the {\em observable} X-ray properties of the theory. The 
results of a spectral fit are a complex function of the the density 
and temperature distributions of the source, absorption, the 
properties of the detector used and the spectral fitting procedure. 
As such they should not be considered as ``real'' values, but as 
characteristic values, and specific to the instrument used. The normal
method of fitting a simplistic model to the observed data, and then treating 
the best-fit parameters as the real properties can easily give answers
an order of magnitude out from the truth. 

This technique will allow the first direct comparison between observation 
and theoretical models of superbubbles and starburst driven outflows.\\

We would like to thank Trevor Ponman, Robin Williams and the referee for 
constructive criticism. DKS and IRS acknowledge financial support from PPARC. 
This work was performed on the Birmingham 
node of the {\sc Starlink} network.

\label{lastpage}
\end{document}